\documentclass[fleqn]{mn}

\usepackage{amsfonts}
\usepackage{amsmath}
\usepackage{graphicx}

\font\sevenrm=cmr7 scaled 1000
\def\gsim{\;\lower4pt\hbox{${\buildrel\displaystyle >\over\sim}$}\;}
\def\lsim{\;\lower4pt\hbox{${\buildrel\displaystyle <\over\sim}$}\;}
\def\grls{\;\lower4pt\hbox{${\buildrel\displaystyle >\over <}$}\;}

\begin{document}

\title[Self-Similar EECC Solutions]
{Envelope Expansion with Core Collapse. \\
I. Spherical Isothermal Similarity Solutions}
\author[Y.-Q. Lou and Y. Shen]
{Yu-Qing Lou$^{1,2,3}$ and Yue Shen$^{1}$ \\
$^1$Physics Department, The Tsinghua Center for Astrophysics,
Tsinghua
University, Beijing 100084, China\\
$^2$Department of Astronomy and Astrophysics,
The University of Chicago,
5640 S. Ellis Ave., Chicago, IL 60637, USA\\
$^3$National Astronomical Observatories, Chinese Academy
of Sciences, A20, Datun Road, Beijing 100012, China.}
\date{Accepted 200?...; Received 2003 ...; in original form 2003 September 19}\maketitle

\begin{abstract}
We investigate self-similar dynamical processes in an isothermal
self-gravitational fluid with spherical symmetry. In reference to
earlier complementary solution results of Larson, Penston, Shu,
Hunter and Whitworth \& Summers, we further explore the
`semi-complete solution space' from an initial instant
$t\rightarrow 0^{+}$ to a final stage $t\rightarrow +\infty$.
These similarity solutions can describe and accommodate physical
processes of radial inflow, core collapse, oscillations and
envelope expansion (namely, outflow or wind) or contraction as
well as shocks. In particular, we present new classes of
self-similar solutions, referred to as `envelope expansion with
core collapse' (EECC) solutions, that are featured by an interior
core collapse and an exterior envelope expansion concurrently. The
interior collapse towards the central core approaches a free-fall
state as the radius $r\rightarrow 0$, while the exterior envelope
expansion gradually approaches a constant radial flow speed as
$r\rightarrow +\infty$. There exists at least one spherical
stagnation surface of zero flow speed that separates the core
collapse and the envelope outflow and that travels outward at
constant speed, either subsonically or supersonically, in a
self-similar manner. Without crossing the sonic critical line
where the travel speed of nonlinear disturbances relative to
the radial flow is equal to the sound speed,
there exist a continuous band of infinitely many EECC solutions
with only one supersonic stagnation point as well as a continuous
band of infinitely many similarity solutions for `envelope
contraction with core collpase' (ECCC) without stagnation point.
Crossing the sonic critical line twice analytically, there exist
infinitely many discrete EECC solutions with one or more subsonic
stagnation points. Such discrete EECC similarity solutions
generally allow radial oscillations in the subsonic region between
the central core collapse and the outer envelope expansion. In
addition, we obtained complementary discrete ECCC similarity
solutions that cross the sonic critical line twice with subsonic
oscillations. In all these discrete solutions, subsonic spherical
stagnation surfaces resulting from similarity oscillations travel
outward at constant yet different speeds in a self-similar manner.
With specified initial boundary or shock conditions, it is possible
to construct an infinite number of such EECC similarity solutions,
which are conceptually applicable to various astrophysical problems
involving gravitational collapses and outflows. We mention potential
applications of EECC similarity solutions to the formation process
of proto-planetary nebulae connecting the AGB phase and the planetary
nebula phase, to H{\sevenrm II} clouds surrounding star formation
regions, and to a certain evolution phase of galaxy clusters.
\end{abstract}

\begin{keywords}
hydrodynamics --- planetary nebulae ---
stars: AGB and post-AGB --- stars: formation
--- stars: winds, outflows --- waves
\end{keywords}

\section{Introduction}

Hydrodynamical processes of a self-gravitational spherical gas
have been investigated from different perspectives and in various
contexts of astrophysical or cosmological problems on entirely
different temporal and spatial scales. As an important example,
solutions to the model
problem of spherical gravitational collapse provide physical
insights for understanding processes of star formation and
outflows (Ebert 1955; Bonner 1956; Hayashi 1966; Hunter 1967;
Spitzer 1968; Larson 1973; Mestel 1974). Another important field
is the formation and evolution of galaxy clusters that contain
massive dark matter haloes and hot gases with temperatures of
$\sim 10^7-10^8$K (e.g. Fabian 1994; Sarazin 1988; Gunn \& Gott
1972; Bertschinger 1985; Fillmore \& Goldreich 1984; Navarro,
Frenk \& White 1996). Detailed numerical computations to solve the
fully coupled nonlinear partial differential hydrodynamic
equations show that solutions, sufficiently away from initial and
boundary conditions, appear to evolve into self-similar behaviours
(e.g. Sedov 1959; Landau \& Lifshtiz 1959). Over past several
decades, researchers have studied self-similar solutions under
spherical symmetry and isothermal conditions. Simultaneously and
independently, Penston (1969a, b) and Larson (1969a, b) found one
discrete self-similar infall solution without involving a central
core mass, the so-called LP-solution referred to in the
literature, on the basis of their numerical calculations of
spherically symmetric gravitational collapse for star formation.
In addition to this LP-solution, Shu (1977) derived other types of
self-similar solutions, including the so-called `expansion-wave
collapse solution' (with a discontinuous derivative at the sonic
critical point), which leads to the physical scenario of
`inside-out collapse' in the context of molecular cloud and/or
star formation (e.g. Shu, Adams \& Lizano 1987). Shu (1977) showed
clearly the existence of the sonic critical line across which two
types of eigensolutions may be possibly specified.

Following Shu's analysis, Hunter (1977) found a class
of discrete `complete solutions' by extending the time domain way
back to the pre-catastrophic period starting from $t\rightarrow
-\infty$ and by applying a zero-speed boundary condition there
(see Fig. 6). One  may regard the LP-solution as a special
member of Hunter's class and Shu's expansion-wave collapse
solution turns out to be a limiting case of Hunter's solution
class [see solution parameters (11) and figs. 1 and 2 of Hunter
(1977)]. In terms of the `semi-complete solution' structure (see
Fig. 6), the `complete' LP-solution and each of `complete'
solutions $a$, $b$, $c$, $d$ and so forth of Hunter's (1977) can
be split into two branches (i.e. one inflow and one outflow) at
$x\equiv r/(at)\rightarrow +\infty$. In the `semi-complete'
perspective, Hunter's solutions $a$, $b$, $c$, $d$ and so forth
involve radial oscillations with increasing number of nodes or
stagnation points in the subsonic region with either envelope
expansion or envelope contraction but without initial onset of
central core collapse (see Fig. 6).

Within the same overall framework of Shu (1977) and Hunter (1977),
Whitworth \& Summers (1985, hereafter WS) introduced more
elaborate mathematical techniques to construct similarity
solutions across the sonic critical line with weak discontinuities
(Lazarus 1981; Hunter 1986; Ori \& Piran 1988; Boily \& Lynden-Bell
1995). Whitworth \& Summers noted the existence of two types of
sonic points (namely, saddle and nodal points defined later) in the
problem (see Jordan \& Smith 1977 for details) and carefully
examined stability properties of numerical integration directions.
They realized the distinctly different properties of Shu's two
types of eigensolutions at the sonic critical line and referred to
the two types as `primary' and `secondary' directions\footnote{For
the two eigensolutions crossing the nodal sonic critical point and
in terms of absolute values, the one of larger flow speed gradient
is along the primary direction and the one of smaller flow speed
gradient is along the secondary direction.} in the case of nodal
points (Jordan \& Smith 1977). By allowing one solution to pass
the nodal sonic point along the secondary direction and an
infinite number of solutions to pass it tangentially along the
primary direction as well as locally linear combinations of the
two eigensolutions, WS constructed two-parameter continuum bands
of solutions in addition to the previous discrete solutions.
Regarding the analysis of WS, Hunter (1986) promptly pointed out
that the solutions of WS involve weak discontinuities across the
sonic critical line in that physical variables and all their first
derivatives in space and time are continuous\footnote{This does
not include those cases when a solution on one side is along the
primary direction and that on the other side is along the
secondary direction.} and derivatives of some higher orders may
also be continuous. Moreover, Hunter (1986) suggested that the
solutions of WS may be unstable against small perturbations as
numerical computations did not show tendency towards any of such
solutions except for the LP-solution and therefore might not be
acceptable physically. Ori \& Piran (1988) proceeded to verify
Hunter's conjecture and derived a necessary stability criterion
for `complete solutions'.

Lynden-Bell \& Lemos (1988) and Lemos \& Lynden-Bell (1989a,
1989b) studied self-similar solutions for a cold fluid, not only
in the Newtonian regime but also in the regime of general
relativity. Foster \& Chevalier (1993) studied the gravitational
collapse of an isothermal sphere by hydrodynamic simulations.
They recovered the LP-solution in the central region where a core
forms and the self-similar solutions of Shu (1977) when the ratio
of initial outer cloud radius to core radius is $\gsim 20$.
Boily \& Lynden-Bell (1995) studied the self-similar solutions
for radial collapse and accretion of radiative gas in the
polytropic approximation. Hanawa \& Nakayama (1997) also
investigated the stability problem of the complete self-similar
solutions in a normal mode analysis and concluded that only the
LP-solution is stable. The most recent work of Harada, Maeda \&
Semelin (2003) studied spherical collapse of an isothermal gas
fluid through numerical simulations and showed a critical
behaviour for the Newtonian collapse that turns out to follow the
self-similar form of the first member (i.e. solution $a$) of
Hunter's solutions (Hunter 1977).

Lai \& Goldreich (2000) studied the growth of nonspherical perturbations
in the collapse of a self-gravitating spherical gas cloud. They found
through numerical computations that nonspherical perturbations damp in
the subsonic region but grow in the supersonic region with asymptotic
scaling relations for their growths. Lai \& Goldreich mentioned potential
applications to core-collapse of supernova explosions (Goldreich \& Weber
1980; Yahil 1983), where the asymmetric density perturbation may lead to
asymmetric shock propagation and breakout, giving rise to asymmetry in
the explosion and a kick to the new-borne neutron star. Subsequently,
Lai (2000) presented a global stability analysis for a self-similar
gravitational collapse of a polytropic gas.

In addition to the shock-free self-similar solutions, Tsai \& Hsu
(1995) found a similarity shock solution in reference to Shu's
expansion-wave collapse solution through both numerical and
analytical analyses for the formation of low-mass stars. Their
results were recently expanded by Shu, Lizano, Galli, Cant\'o \&
Laughlin (2002) in the context of self-similar `champagne flows'
driven by a shock in H{\sevenrm II} regions. One shock solution found by
Tsai \& Hsu (1995) turns out to be the limiting case of many
self-similar shocked LP-solutions found by Shu et al. (2002).

The perspectives of complete (Hunter 1977, 1986) and semi-complete
(Shu 1977; WS) similarity solutions are both valid with proper
physical interpretations and with the corresponding identification
of an initial moment. In view of the invariance under the
time-reversal transformation of self-gravitational fluid equations
$(1)-(3)$ below, that is,
$t\rightarrow -t$, $\rho\rightarrow\rho$, $u\rightarrow -u$,
the complete and semi-complete similarity solutions are
closely related to each other. For example, the complete
LP-solution (fig. 1 of Hunter 1977) describes a self-similar
collapse process starting from the pre-catastrophic instant
$t\rightarrow-\infty$ with a zero flow speed and a density
distribution, passing through the sonic critical point once,
gaining a finite radial infall speed at $t=0$, and approaching
eventually a free-fall state involving a central core collapse
as $t\rightarrow +\infty$. However, the complete LP-solution
(Hunter 1977) can be broken into two branches of similarity
solution in the semi-complete space. One branch is only
partially shown in the lower-left corner of Shu's fig. 2 and
is described by equations $(A7)$ and $(A8)$ in Shu's appendix
(see also Fig. 5). This branch corresponds to a self-similar
outflow without core mass but with a finite central density
decreasing with time $t$ in a scaling of $t^{-2}$. The other
branch corresponds a constant radial inflow speed at
$x\rightarrow +\infty$ and a core collapse at
$x\rightarrow 0^{+}$ without crossing the sonic critical line.
To some extent, Shu's criticism on the utility of the semi-complete
LP-solution may be compromised by joining the other branch of
the semi-complete LP-solution of infall-core-collapse solution
(Hunter 1977) or by inserting a shock to match with an outward
`breeze' for `champagne flows' in H{\sevenrm II} regions (Tsai
\& Hsu 1995; Shu et al. 2002).

With the complementarity of complete and semi-complete
perspectives for self-similar solutions, we re-visit this classic
problem in the semi-complete solution space with the strong
motivation of finding similarity solutions for concurrent
processes of central core collapse and envelope expansion or
contraction that may or may not involve subsonic radial
oscillations and/or shocks across the sonic critical line.
Specifically, we have learned Shu's expansion-wave collapse
solution as the limit of a class of core collapse solutions with
zero infall speed at $x\rightarrow +\infty$. We have also noted
that Hunter's complete collapse solutions, including the
LP-solution, can be used to describe radial outflows with or
without subsonic radial oscillations in the semi-complete
perspective. It is then physically plausible to derive
self-similar solutions for concurrent envelope expansion with core
collapse (EECC) that may or may not involve subsonic radial
oscillations and/or shocks. If one allows solutions that pass
through the sonic critical line with weak discontinuities (Lazarus
1981), then the mathematical solution space can be further
expanded enormously (WS). This paper gives a positive account
of constructing EECC
similarity solutions that are distinctly different from known
similarity solutions. With proper selections and adaptations, we
believe that these EECC similarity solutions, due to their
generality and simplicity, provide an important basic conceptual
framework for a wide range of astrophysical problems involving
gravitational collapses, outflows, contractions and subsonic
radial oscillations etc. We shall mention a few potential
applications, including the dynamical process of forming a
proto-planetary nebula that evolves from the asymptotic giant
branch (AGB) phase to the planetary nebula phase with a central
hot proto white dwarf. Specifics will be described in forthcoming
papers.

%

Several authors have approached the same set of basic isothermal
fluid equations with different notations, choice of variables and
conventions of presenting their results. It is sometimes
mind-boggling to find correspondences among their solutions and
results of analysis. Without further complications, we shall more
or less adopt Shu's notations and present our results in a similar
manner. The basic equations and the self-similar transformation
are summarized in Section 2. In contrast to Shu's focus on the
first quadrant of $-v$ versus $x$, we pay equal attention to both
first and fourth quadrants for $-v$ versus $x$ presentation as we
are interested in both outflows and collapses. In order to place
our work in a proper perspective in reference to known results,
we have explored the entire solution structure, including the
counterparts of those solutions found by previous authors, in
Section 3. Physical interpretations for the obtained solutions are
described in Section 4. We summarize and discuss our results in
Section 5. Some technical details and procedures can be found in
the Appendix.

\section{The Basic Model Formulation}

To be self-contained, we recount basic isothermal gravitational
hydrodynamic equations with spherical symmetry below in the
spherical polar coordinate ($r,\theta,\varphi$). We cast these
nonlinear equations in a self-similar form (Larson 1969a; Penston
1969a; Shu 1977; Hunter 1977, 1986; Whitworth \& Summers 1985) to
derive useful conditions and solutions. By the spherical symmetry,
the flow velocity is purely radial and the mass conservation
equation is
\begin{equation}
\frac{\partial\rho}{\partial t}
+\frac{1}{r^2}\frac{\partial}{\partial r}(r^2\rho u)=0\ ,
\end{equation}
where $\rho$ is the gas mass density, $u$ is the bulk radial flow
speed and $t$ is the time.

Equivalently, it is informative to rewrite equation (1) in terms
of mass enclosed in a spherical shell, namely
\begin{equation}
\frac{\partial M}{\partial t}+u\frac{\partial M}{\partial r}=0\ ,\
\qquad\qquad\qquad\frac{\partial M}{\partial r}=4\pi r^2\rho\ ,
\end{equation}
where $M(r,t)$ is the total mass enclosed
inside a sphere of radius $r$ at time $t$.

In the isothermal approximation with $p=a^2\rho$, where $p$ is the
gas pressure and $a$ is the isothermal sound speed, the radial
momentum equation then reads
\begin{equation}
\frac{\partial u}{\partial t}+u\frac{\partial u}
{\partial r}=-\frac{a^2}{\rho}\frac{\partial\rho}
{\partial r}-\frac{GM}{r^2}\ ,
\end{equation}
where $G\equiv 6.67\times 10^{-8}\hbox{ dyne cm}^2\hbox{ g}^{-2}$
is the gravitational constant.

In equation (3), one identifies
$-\partial\Phi/\partial r\equiv -GM(r, t)/r^2 ,$
where $\Phi(r,t)$ is the gravitational potential. In this form,
the Poisson equation for the gravitational potential $\Phi$ is
automatically satisfied. The equation of energy conservation can
be readily derived by combining equations $(1)-(3)$, namely
\begin{equation}
\begin{split}
\frac{\partial}{\partial t}\bigg\lbrace\frac{\rho u^2}{ 2}
+a^2\rho \bigg[\ln\bigg(\frac{\rho}{\rho_c}\bigg)-1\bigg]
-\frac{1}{8\pi G}
\bigg(\frac{\partial\Phi}{\partial r}\bigg)^2\bigg\rbrace\\
+\frac{1}{r^2}\frac{\partial}{\partial r} \bigg\lbrace
r^2u\rho\bigg[\frac{u^2}{2}
+a^2\ln\bigg(\frac{\rho}{\rho_c}\bigg)\bigg]\bigg\rbrace\\
+\frac{1}{r^2}\frac{\partial}{\partial r}(r^2u\rho\Phi)
+\frac{1}{r^2}\frac{\partial}{\partial r} \bigg(\frac{r^2\Phi}{
4\pi G}\frac{\partial^2\Phi}{\partial r\partial t}\bigg)=0\
\end{split}
\end{equation}
(Fan \& Lou 1999), where $\rho_c$ is an arbitrary constant density
scale by the mass conservation (1). One can clearly separate out
two terms associated with $\rho_c$ in the energy conservation
equation (4), with the meaning of energy assigned to gas particles.
For the present problem, it turns out that the last two divergence
terms related to $\Phi$ on the left-hand side of equation (4) exactly
cancel each other using the two expressions of equation (2). By
equation (4), it is straightforward to identify the energy density
${\cal E}$ and energy flux density ${\cal F}$, respectively [see
definition equations (23) and (24) later].

If one were to derive similarity `shock conditions' across a spatial
discontinuity, the joint requirements of mass conservation (1),
momentum conservation (3) and energy conservation (4) together would
strictly imply continuous solutions only. That is, no simultaneous
jumps in density, radial flow velocity and temperature are allowed
across a spatial point in a similarity form. However in astrophysical
situations (such as H{\sevenrm II} regions in molecular clouds) where
radiative cooling is sufficiently fast at the shock location and the
movement of shock front is sufficiently slow, one may construct
`isothermal shocks' (e.g. Spitzer 1978) based on mass conservation (1)
and momentum conservation (3) (e.g. Courant \& Friedrichs 1948) and
leave radiative losses to accommodate energy conservation in a form
different from equation (4).

To derive self-similar solutions of equations $(1)-(3)$, we
introduce the dimensionless independent self-similar variable
$x=r/at$. Dependent variables then take the following similarity
forms accordingly, namely
\begin{eqnarray}
\rho(r,t)=\frac{\alpha(x)}{4\pi Gt^2}\ ,\
\qquad\quad M(r,t)=\frac{a^3t}{G}m(x)\ ,\nonumber
\end{eqnarray}
\begin{equation}
u(r,t)=av(x)\ ,\qquad\qquad \Phi(r,t)=a^2\phi(x)\ ,
\end{equation}
where dimensionless $\alpha$, $m$, $v$ and $\phi$ are reduced
variables for density, enclosed mass, radial flow speed and
gravitational potential, respectively; they are functions of $x$
only. In our formulation, $t=0$ specifies the initial instant.
Our choice of similarity variables is the same as that of Shu
(1977) and of WS, while other authors may adopt different yet
equivalent forms that are of the same physical nature.

Substitution of transformation equation (5) into equation (2) yields
\begin{equation}
m+(v-x)\frac{dm}{dx}=0\ ,
\qquad\qquad
\frac{dm}{dx}=x^2\alpha\ .
\end{equation}
It follows immediately that
\begin{equation}
m=x^2\alpha(x-v)\ ,
\end{equation}
which leads to
\begin{equation}
\frac{d}{dx}[x^2\alpha(x-v)]=x^2\alpha\ .
\end{equation}
For $x>0$, the physical requirement that $m(x)>0$ means $x-v>0$
by equation (7). Therefore, portions of similarity solutions of
$-v(x)$ and $\alpha(x)$ to the lower-left of straight line $x-v=0$
(parallel to the sonic critical line $x-v=1$) are unphysical.

Substitution of transformation equation (5) into equations (1)
and (3) together with relation (7) yields a pair of two
coupled nonlinear ordinary differential equations (ODEs), namely
\begin{equation}
[(x-v)^2-1]\frac{dv}{dx}
=\bigg[\alpha(x-v)-\frac{2}{x}\bigg](x-v)\ ,
\end{equation}
\begin{equation}
[(x-v)^2-1]\frac{1}{\alpha}\frac{d\alpha}{dx}
=\bigg[\alpha-\frac{2}{x}(x-v)\bigg](x-v)\
\end{equation}
[see equations (11) and (12) of Shu 1977]. It is then easy to show
that equation (8) can be derived from equations (9) and (10)
without involving the sonic critical point at $(x-v)^2=1$. The
main thrust of the present work is to solve the above coupled pair
of nonlinear ODEs and search for new EECC self-similar solutions.

It is also straightforward to derive the Bernoulli relation
along streamlines from equations $(1)-(3)$, or equivalently,
from nonlinear ODEs (9) and (10), namely
\begin{equation}
f(x)-x\frac{df}{dx}+\frac{v(x)^2}{2}
+\ln\alpha(x)+\phi(x)=\hbox{constant}\ ,
\end{equation}
where $f(x)$ and $\phi(x)$ are functions of $x$
only and satisfy the following two relations
\begin{equation}
\frac{df}{dx}=v(x)\ \qquad\hbox{ and }\qquad
\frac{d\phi}{dx}=\frac{m(x)}{x^2}=\alpha(x-v)\ ,
\end{equation}
respectively.
Since $m(x)>0$ for $x>0$, the reduced gravitational potential
$\phi(x)$ is a monotonically increasing function of $x$.

\section{Self-similar solutions of isothermal gas flows}

All information of self-gravitational spherical isothermal
similarity flows is contained in the nonlinear ODEs summarized in
the preceding section. Our primary focus is to derive and analyze
possible similarity solutions of nonlinear ODEs (9) and (10) and
their properties. In general, these similarity solutions can
describe radial outflows (winds), inflows (contractions),
oscillations, collapses (infalls) with or without shocks
(including weak discontinuities etc.) and with or without central
singularities at $x\rightarrow 0^{+}$. Before proceeding, it would be
important to clarify several relevant concepts. We search for
continuous solutions of both reduced velocity $v(x)$ and reduced
density $\alpha(x)$ as functions of the independent similarity
variable $x$; the reduced enclosed mass $m(x)$ and the reduced
gravitational potential $\phi(x)$, as well as other pertinent
physical variables can be obtained accordingly.

In this paper, we shall take the range of the independent
similarity variable $x$ to extend from $x\rightarrow+\infty$ to
the origin $x\rightarrow 0^{+}$. Such similarity solutions are
referred to as `semi-complete solutions' (WS) in contrast to the
`complete solutions' (Hunter 1977) that range from
$x\rightarrow-\infty$ to $x\rightarrow+\infty$. Since $x\equiv
r/(at)$, at a given instant $t>0$, we have the entire radial range
of $r$, while at a given radial location $r$, we have the entire
temporal range of $t>0$. For example, the range of
$x\rightarrow+\infty$ to $x=0$ may correspond the initial instant
of $t=0$ when the core begins to collapse to the final epoch
$t\rightarrow+\infty$. On the other hand, the `complete solutions'
(Hunter 1977) extend back to the pre-catastrophic era from
$t\rightarrow-\infty$ to $t=0$. One major advantage of considering
`semi-complete solutions' is that the boundary condition of zero
flow speed at $t\rightarrow-\infty$ becomes unnecessary. It is
then possible to construct more physical solutions of interest
with various initial conditions at $t\rightarrow 0^{+}$ and
boundary conditions at $r\rightarrow 0$. As the relevant nonlinear
ODEs here can be satisfied under the time-reversal transformation:
$x\rightarrow-x$, $v\rightarrow-v$, $\alpha\rightarrow\alpha$ and
$m\rightarrow-m$, the `complete solutions' and `semi-complete
solutions' relate to each other as will be discussed later.
Moreover, the time reversal transformation can also lead to
different interpretations and/or applications of self-similar
solutions.

\subsection{Special Analytical Solutions}

A straightforward exact analytical solution of
nonlinear ODEs (9) and (10) is the similarity solution
\begin{equation}
v=0\ ,\quad \alpha=\frac{2}{x^2}\ ,\quad m=2x\ , \quad
\frac{d\phi}{dx}=\frac{2}{x}\ ,
\end{equation}
which describes a hydrostatic state of a spherical system (Ebert
1955; Bonner 1956; Chandrasekhar 1957; Shu 1977). Note that equation
(10) with $v=0$ gives a solution of $\alpha=2/(Cx^2+1-C)$ where $C$
is an integration constant. Independently, according to equations
(7) and (8) with $v=0$, one derives $\alpha=C'/x^2$ where $C'$ is
another integration constant. The two integration constants $C$
and $C'$ can be chosen (i.e. $C=1$ and $C'=2$) such that the two
$\alpha(x)$ solutions are consistent with each other. For a composite
spherical flow system of two isothermal fluids under self-gravity,
such type of similarity solutions does not exist, even though a
hydrostatic solution in a composite system does exist (Lou et al.
2003, in preparation).

Another exact analytical similarity solution
of nonlinear ODEs (9) and (10) is
\begin{equation}
v=\frac{2}{3}x\ ,\quad \alpha=\frac{2}{3}\ , \quad
m=\frac{2}{9}x^3\ ,\quad \frac{d\phi}{dx}=\frac{2}{9}x\ ,
\end{equation}
(WS) that passes the sonic critical line at $x_*=3$ and that
represents an expanding outflow with $v$ approaches infinity
as $x\rightarrow+\infty$. Such type of exact solutions also
exist in a composite system of two isothermal fluid spheres
coupled by self-gravity (Lou et al. 2003, in preparation).
It is interesting to note that this solution actually corresponds
to a nonrelativistic Hubble expansion in the model of the
Einstein-de Sitter universe (e.g. Shu et al. 2002). In dimensional
form, a pertinent Hubble `constant' here is $H=2/(3t)$ and a
critical mass density is $\rho=3H^2/(8\pi G)=1/(6\pi Gt^2)$.

Another explicit solution is the singular solution
\begin{equation}
x-v=1\ ,\qquad\qquad \alpha(x)=2/x \ .
\end{equation}
[The conjugate set of $x-v=-1$ and $\alpha(x)=-2/x$
is not considered here due to the requirement of
$\alpha(x)\geq 0$. The situation would be different in
a time reversal problem; see Hunter 1977]. Apparently,
such a singular solution does not satisfy equation (8)
and is not a real solution to the problem. It does
set a requirement for sensible solutions that pass
smoothly across the sonic critical line at $x-v=1$.
We will describe below more specifically the way of passing
through a sonic critical point smoothly.

\subsection{Asymptotic Solution Behaviours}

Except for the non-relativistic Einstein-de Sitter solution (14), a
reasonable assumption is that when $x$ goes to infinity the reduced
radial velocity $v(x)$ should approach a finite value\footnote{In
Shu (1977), the condition that the reduced radial velocity $v(x)$
vanishes as $x\rightarrow+\infty$ was specified for his collapse
solutions (without crossing the sonic critical line) with the
`expansion-wave collapse' as the limiting solution.}. Such
similarity solutions can describe either outflows with $v(x)>0$
or inflows with $v(x)<0$ or quasi-static or `breeze' state
$v(x)\rightarrow 0$ at large radii (i.e. $x\rightarrow +\infty$).

One readily derives from nonlinear ODEs (9)
and (10) asymptotic solutions at large radii
\begin{equation}
\begin{split}
&v=V+\frac{2-A}{x}+\frac{V}{x^2}
+\frac{(A/6-1)(A-2)+2V^2/3}{x^3}+\cdots
\ ,\\
&\alpha=\frac{A}{x^2}+\frac{A(2-A)}{2x^4}
+\frac{(4-A)VA}{3x^5}+\cdots\ ,
\end{split}
\end{equation}
where $V$ and $A$ are two independent constants.\footnote{Note
that asymptotic solution (3.8) of WS contains a typo, that is,
their $4y_{\infty}$ should be replaced by $4y_{\infty}^2$.}
These asymptotic solutions are more general than those of
Shu (1977) which can be recovered by simply setting $V=0$.

Similarly, we derive two types of asymptotic solutions
when approaching the origin $x=0$, either
\begin{equation}
v\rightarrow-(2m_0/x)^{1/2},
\ \ \alpha\rightarrow[m_0/(2x^3)]^{1/2},\
 \  m\rightarrow m_0\ ,
\end{equation}
or
\begin{equation}
v\rightarrow\frac{2}{3}x
\ ,\quad \
\alpha\rightarrow B
\ ,\quad
\ m\rightarrow Bx^3/3\ ,
\end{equation}
where $m_0$ and $B$ are two constants (WS). The first asymptotic
solution (17) for $x\rightarrow 0^{+}$ describes a central
free-fall state during a self-similar collapse, with a
finite core mass that increases with time and a diverging
gravitational potential that steepens with increasing time. The
second asymptotic solution (18) for $x\rightarrow 0^{+}$ does not
involve central core mass and has a finite gravitational potential
that becomes shallower with increasing time. Both types of
$x\rightarrow 0^{+}$ solutions can be matched with outflow or
contraction solutions at large $x$ with or without subsonic radial
oscillations. Solution (18) also corresponds to an infall or collapse
in a time reversal problem. We note that solution (18) with
$x\rightarrow 0^{-}$ is in fact the no-flow boundary condition as
$t\rightarrow -\infty$ for the `complete solutions' of Hunter
(1977) and subsequent papers.



Asymptotic solutions $(16)-(18)$ are extremely valuable for
estimating compatible values of $v(x)$ and $\alpha(x)$ when
performing numerical integrations to obtain similarity
solutions starting from either $x\rightarrow +\infty$ or
$x\rightarrow 0^{+}$.

We now consider solution behaviours in the vicinity of the
sonic critical line. Except for inserting shocks across the
sonic critical line to match two different branches of
isothermal similarity solutions (Tsai \& Hsu 1995; Shu et al.
2002), similarity solutions are physically acceptable when
condition (15) is met across the sonic critical point at
$x-v=1$. As already noted (Shu 1977; Hunter 1977, 1986; WS),
the sonic critical line separates subsonic regions, where the
flow is subsonic relative to the similarity profile, from
supersonic regions, where the flow is supersonic. Two types
of sonic points can be identified. In our notations, the sonic
points with $0<x_*<1$ in the first quadrant are all saddle
points, while those with $x_*>1$ in the fourth quadrant are
all nodal points (Jordan \& Smith 1977).

In general, there exist two types of eigensolutions at one sonic
critical point with their first derivatives being computed by
L'H\^{o}pital's rule\footnote{Higher-order derivatives at a
sonic critical point can be readily computed.
We simply list derivatives up to the third order in the
Appendix. Despite typos [e.g. the $z_s^{\prime}$ in equation
(3.3) of WS should be $\pm 2/x_s^2$] and different notations,
the calculated derivatives here are the same as those obtained
by previous authors.}, namely
\begin{equation}
\frac{dv}{dx}\bigg|_{x=x_*}=1-\frac{1}{x_*}\ ,
\qquad
\frac{d\alpha}{dx}\bigg|_{x=x_*}
=-\frac{2}{x_*}\bigg(\frac{3}{x_*}-1\bigg)
\end{equation}
for the type 1 eigensolution and
\begin{equation}
\frac{dv}{dx}\bigg|_{x=x_*}=\frac{1}{x_*}\ ,
\qquad\qquad
\frac{d\alpha}{dx}\bigg|_{x=x_*}=-\frac{2}{x_*^2}
\end{equation}
for the type 2 eigensolution. Our classification here follows that
of Shu (1977), Hunter (1977) and WS. It is then clear that the two
types of derivatives of $v(x)$ are of the opposite signs for a
saddle point ($0<x_*<1$) and are both positive for a nodal point
($x_*>1$); they are positive and equal at $x_*=2$ and their
magnitudes reverse for $x_*>2$. Mathematically, when the sonic
point is a saddle, only eigensolutions can go across the sonic
critical point; when the sonic point is a node, besides the two
eigensolutions there are infinitely many solutions which can go
across a nodal point tangential to the direction of the
eigensolution with a larger gradient: this eigendirection is
referred to as the primary direction, while the other is referred
to as the secondary direction that allows only one solution to
pass through. The two eigensolutions are analytic across the sonic
critical point (Hunter 1977, 1986; Ori \& Piran 1988). Other
solutions passing through a nodal sonic point along the primary
direction involve weak discontinuities (Lazarus 1981; WS; Hunter
1986; Ori \& Piran 1988; Boily \& Lynden-Bell 1995). Specific
properties of the sonic point are detailed in Appendix A.

\subsection{Numerical Solutions}

We now explore and present the similarity solution structure from
the perspective of `semi-complete solutions' in reference to
previously known solutions (Larson 1969a; Penston 1969a; Shu 1977;
Hunter 1977, 1986; Whitworth \& Summers 1985). As already noted
earlier and will be further discussed below, there exists a
correspondence between the `complete' and `semi-complete'
solutions. In the following, we construct numerical solutions in
the range from $x\rightarrow+\infty$ to $x\rightarrow 0^{+}$ by
making use of various asymptotic similarity solutions (see
subsections 3.1 and 3.2). When $x$ approaches infinity, solutions
should converge to asymptotic solution (16), while when $x$
approaches the origin, solutions should match with either
solution (17) or solution (18). Among others, those solutions
without encountering the sonic critical line can be readily
obtained by numerical integrations from the asymptotic solution
(16) at large $x$. However, for those similarity solutions
crossing the sonic critical line, we must take extra care to
distinguish various types of similarity solutions and their
analyticity across the sonic critical line.

\subsubsection{Solutions without crossing the sonic critical line}

It is straightforward to derive numerically those similarity
solutions in the entire $x$ range without crossing the sonic
critical line. We refer to the two constant parameters $V$
and $A$ in asymptotic solution (16) as asymptotic radial flow
speed and mass density parameters, respectively. Starting from
asymptotic solution (16) for $x\rightarrow+\infty$ with the two
parameters $V$ and $A$ specified, we use the standard fourth-order
Runge-Kutta scheme to integrate nonlinear ODEs (9) and (10) from
a sufficiently large $x$ to small $x$. More general than the
condition that $v(x)$ vanishes at $x\rightarrow+\infty$ (e.g. Shu
1977), parameter $V$ in asymptotic solution (16) is allowed here
to take either positive (outflows) or negative (inflows) values.

\begin{figure}
\begin{center}
\includegraphics[scale=0.65]{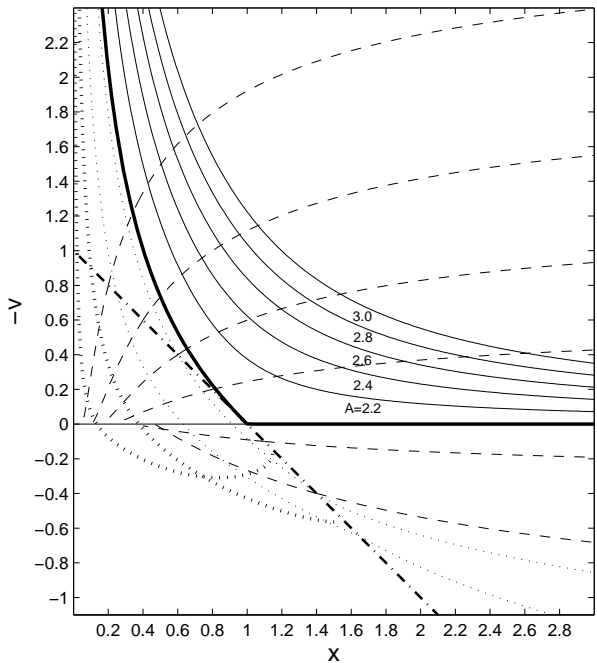}
\caption{Representative examples of similarity solutions shown as
$-v(x)$ versus $x$. Solutions without crossing the sonic critical
line, the straight dash-dotted line for $x-v=1$, for $V=0$ and
$A>2$ are plotted in light solid curves, with the heavy solid
curve for the `expansion-wave collapse' of $A\rightarrow 2^{+}$
as the limiting solution. In both first and fourth quadrants, the
type 1 solutions passing across the sonic critical line are plotted
in dashed curves and are artificially cut off at the $x$ axis. The
special type 1 solution with $x_*=1$ is simply static solution (13).
The type 2 solutions expanded in the first quadrant with critical
points are shown by dotted lines; some can go smoothly across the
sonic critical line again in the fourth quadrant and extend to the
right infinity, although they may not be analytic (i.e. with weak
discontinuities) at the critical point in the fourth quadrant (WS;
Hunter 1986). It should be noted that not all dotted lines can pass
across the sonic critical line twice.
We plot two such `unfinished' type 2 solutions with first critical
points at $x_*=0.1$ and $x_*=0.02$ in heavy dotted-lines as
illustrating examples. Behaviours of both type 1 and type 2
solutions as $x\rightarrow+\infty$ are determined by solution
(16). Some of the corresponding reduced density curves are shown
in Fig. 4.}
\end{center}
\end{figure}

For comparison and for setting up the stage, we reproduce Shu's
light solid line solutions (see fig. 2 of Shu 1977) by setting $V=0$
together with different values of $A>2$ and also plot them in light
solid curves in Fig. 1 as references. Such solutions have the
asymptotic behaviours of solution (16) at large $x$ and of solution
(17) near the origin $x=0$. These solutions with $A>2$ of Shu (1977)
represent self-similar collapse inflows everywhere with the fluid
being at rest at infinity [$x\equiv r/(at)\rightarrow+\infty$].
Naturally, the value of the reduced
core mass parameter $m_0$ is determined by the value of $A$ or
vice versa. The limiting solution of $V=0$ and $A\rightarrow 2^+$
is of considerable interest in the context of star formation (e.g.
Shu et al. 1987) and is shown in Fig. 1 by the heavy solid curve.
This same limiting solution can also be obtained by taking the
limit of $x_*\rightarrow 1^{-}$ where $x_*\le 1$ is the sonic
critical point in the first quadrant. This limiting solution of
$-v(x)$ intersects at $x_{*}=1$ where $-v=0$, $d(-v)/dx=-1$ and joins
the static singular isothermal sphere solution (13) at $x_*=1$
with the same reduced mass density $\alpha$ there. Shu (1977)
referred to this special solution as the {\it expansion-wave
collapse solution} that describes an expansion-wave front outside
which the fluid is at rest and inside which the fluid collapses
to form a central core. This expansion-wave front at
$x\equiv r/(at)=1$ travels outward at the isothermal sound speed
$a$.

Shu's `plus solutions' [dashed curves with positive $d(-v)/dx$
in the first quadrant] pass through the sonic critical line at
$x-v=1$ of his $(A1)$ type (also referred to as type 1 in the
literature) and approach constant speeds at large radii
($x\rightarrow+\infty$). At large $x$, the reduced mass density
scales as $\alpha(x)\ \propto x^{-2}$. At smaller $x\neq 0$ where
$-v$ may becomes subsonic, these `plus solutions' might be matched
with approximate quasi-hydrostatic solutions as indicated by
equations $(A4)-(A6)$ and discussions in the Appendix of Shu (1977).
Instead of time-reversed self-gravitating winds, it would be valid
to regard these `plus solutions' as self-similar inflows with
constant speeds at large $x$. In addition to quasi-static
matchings, it might be possible to match these self-similar inflow
solutions with self-similar outflows at smaller $x$ with or
without shocks.

Shu's `minus solutions' [dotted curves with negative $d(-v)/dx$
in the first quadrant] pass through the sonic critical line at
$x-v=1$ of his $(A2)$ type (also referred to as type 2 in the
literature) and approach zero at $x<1$. As these solutions
join the static singular isothermal sphere solution (13) at $x<1$
with smaller values of $\alpha$, Shu (1977) ignored them but
focused on the limiting expansion-wave collapse solution at
$x_{*}\rightarrow 1^{-}$ where $\alpha$ can be made continuous.
In the present analysis, we emphasize that these similarity
solutions can be further continued in the fourth quadrant to pass
through the second sonic critical point. Those solutions that pass
through the sonic critical line in the fourth quadrant along the
primary direction may involve weak discontinuities (Lazarus 1981;
WS; Hunter 1986). There exists one solution (intersecting the
$x-$axis once)
that passes through the sonic critical line analytically in the
fourth quadrant along the secondary direction. These similarity
solutions are characterized by core collapse infalls for small $x$
and envelope expansions for large $x$ with a stagnation point at
$x<1$ that travels outward at a subsonic speed. These similarity
solutions are referred to as EECC solutions (defined earlier) and
will be further  discussed in the context of forming planetary
nebulae with central proto white dwarfs.

Only a small segment of the LP-solution
is shown in the lower-left corner of fig. 2 of Shu (1977)
with basic properties of LP-solution described in the paragraph
containing equations $(A7)$ and $(A8)$. Hunter (1977) obtained a class
of discrete yet infinitely many LP-type solutions in the `complete'
perspective. These solutions will be discussed and compared with our
`semi-complete' solutions later.


For numerical integrations with $V=0$, condition $A>2$ is
necessary for self-similar solutions to exist without crossing the
sonic critical line $x-v=1$ in the first quadrant. For $V<0$
corresponding to inflow at $x\rightarrow+\infty$, the lower limit
of $A$ such that similarity solutions exist without crossing the
sonic critical line becomes less than $2$. As illustrating
examples shown in Fig. 2, we obtained solutions through numerical
integrations from large $x$ with $V=-2$ and different values of
$A$. When $A$ becomes sufficiently small, similarity solutions
will `crush' towards the sonic critical line (e.g. the two curves
labelled by $A=0.2$ and $A=0$). This `crush' behaviour may be
avoided when $A$ becomes sufficiently large (e.g. $A>0.435$; the
more precise separation value of $A$ lies somewhere between
$A=0.2$ and $A=0.435$). For solutions `crushing' towards the sonic
critical line in Fig. 2, only those satisfying critical condition
(15) in the first quadrant (saddle points) are physically valid.
\begin{figure}
\begin{center}
\includegraphics[scale=0.45]{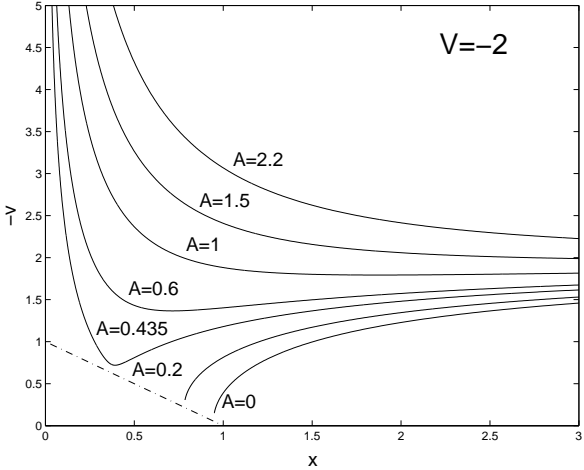}
\caption{The class of similarity solutions integrated from a
sufficiently large $x$ for $V=-2$ with different values of $A$.
For similarity solutions without crossing the sonic critical
line, the dash-dotted line for $x-v=1$, the lower limit of $A$
is less than 2. For $A$ less than this lower limit, solutions
will bump into the sonic critical line; only those satisfying
critical condition (15) are relevant.}
\end{center}
\end{figure}
\begin{figure}
\begin{center}
\includegraphics[scale=0.45]{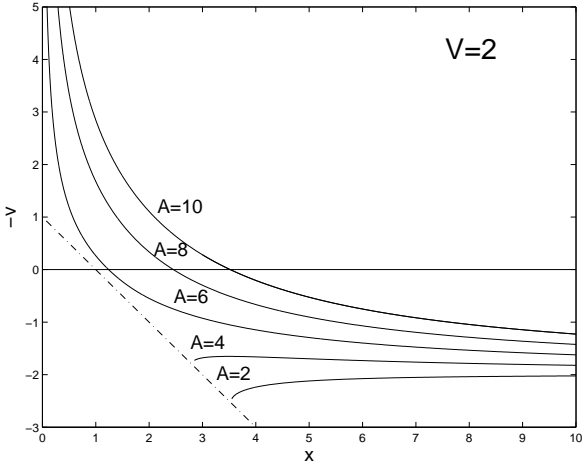}
\caption{The class of similarity solutions integrated from a
sufficiently large $x$ for $V=2$ with different values of $A$.
For similarity solutions without crossing the sonic critical
line, the dash-dotted line for $x-v=1$, the lower limit of $A$ is
greater than 2. For $A$ less than this lower limit, solutions
will bump into the sonic critical line; only those satisfying
critical condition (15) are relevant.}
\end{center}
\end{figure}

Similarly, parameter $V$ can be chosen to be positive corresponding
to a constant outflow speed as $x\rightarrow+\infty$. For different
values of $A$, it is straightforward to integrate numerically from
a sufficiently large $x$ with asymptotic solution (16) towards the
origin $x=0$. The minimum $A$ value for similarity solutions without
crossing the sonic critical line now becomes greater than 2 in
general, and varies for different values of $V>0$. In Fig. 3, we show
examples of $V=2$ with different values of $A$. We emphasize that the
three upper similarity solutions (i.e. $A=6, 8, 10$) describe envelope
expansions with core collapses (EECC) with the stagnation point
travelling outward at a supersonic speed. The reduced core mass
parameter $m_0$ depends upon the two parameters $V$ and $A$. There
exist continuous bands of such EECC similarity solutions in terms of
the two parameters $V$ and $A$.

We have just shown the family of similarity solutions without
crossing the sonic critical line. The class of Shu's solutions is a
special case with parameter $V=0$ in our scheme of classification.
All these similarity solutions approach asymptotic condition (17) as
$x\rightarrow 0^{+}$ with different values of $m_0$ that depends on
both $V$ and $A$. Typically, we find for the same value of $V$, a
larger $A$ will give a larger $m_0$; and for the same value of $A$,
a larger $V$ (including its sign) leads to a smaller $m_0$. This can
be understood as the initial density is proportional to $A$ and the
initial radial velocity is specified by $V$. Inflows and outflows
correspond to $V<0$ and $V>0$, respectively. We provide physical
interpretations for these similarity solutions in Section 4.
Examples of different $V$, $A$ and $m_0$ are summarized in Table 1.

In the radiative gas flow problem considered by Boily \& Lynden-Bell
(1995), a table is listed in their subsection 8.2 but without explicit
solution figures for us to compare. Qualitatively, it appears that
their results in subsection 8.2 might have some similar properties
as our solutions discussed in this subsection, which do not involve
the sonic critical line.

\subsubsection{Solutions crossing the sonic critical line}

Technically, one cannot directly obtain similarity solutions
that go across the sonic critical line by numerical integration
from $x\rightarrow+\infty$. For solutions satisfying critical
condition (15), we expand $v(x)$ and $\alpha(x)$ in Taylor series
in the vicinity of a sonic critical point denoted by $x_*$. For
eigensolutions (e.g. Jordan \& Smith 1977), we have Taylor
expansions near a critical point $x_*$ as
\begin{equation}
\begin{split}
-v(x)=&(1-x_*)+\bigg(\frac{1}{x_*}-1\bigg)(x-x_*)\\
&\qquad\qquad+\frac{x_*-1}{2x_*^2}(x-x_*)^2+\cdots\ ,\\
\alpha(x)=&\frac{2}{x_*}-\frac{2}{x_*}
\bigg(\frac{3}{x_*}-1\bigg)(x-x_*)\\
&\qquad\qquad+\frac{x_*^2-8x_*+13}{x_*^3}(x-x_*)^2+\cdots\
\end{split}
\end{equation}
for type 1 solutions, and
\begin{equation}
\begin{split}
-v(x)=&(1-x_*)-\frac{1}{x_*}(x-x_*)\\
&\qquad\qquad -\frac{x_*^2-5x_*+5}{2x_*^2(2x_*-3)}(x-x_*)^2+\cdots\ ,\\
\alpha(x)=&\frac{2}{x_*}-\frac{2}{x_*^2}(x-x_*)\\
&\qquad\qquad -\frac{x_*^2-6x_*+7}{x_*^3(2x_*-3)}(x-x_*)^2+\cdots\
\end{split}
\end{equation}
for type 2 solutions (Shu 1977; Hunter 1977, 1986; Whitworth \&
Summers 1985). The segment of the sonic critical line in the range
of $0<x_*<1$ in the first quadrant involves saddle points; the type
1 solutions correspond to negative first-order derivatives of $v(x)$,
while the type 2 solutions correspond to positive first-order
derivatives of $v(x)$ (e.g. dotted curves in the first quadrant
of Fig. 1). The portion of the sonic critical line in the range
of $x_*>1$ involves nodal points and both types of eigensolutions
have positive first-order derivatives of $v(x)$ across the sonic
critical line.

It is straightforward to calculate higher-order derivatives
at a critical point $x_*$ (see equations $A2$ and $A3$ in
Appendix A). With these in mind, we perform Taylor expansions
in the vicinity of a sonic critical point to assign compatible
initial values of $v(x)$ and
$\alpha(x)$ and integrate along numerically stable directions. Both
WS and Hunter (1986) noted that numerical integrations away from a
saddle point are stable, while numerical integrations away from a
nodal point are either unstable or neutrally stable (Jordan \&
Smith 1987)\footnote{Numerical integrations towards a nodal point
are stable, while numerical integrations away from a nodal point
along the direction of eigensolutions with the smaller magnitude of
the first derivative (i.e. the secondary direction) are neutrally
stable (WS).}.
We have been extremely careful in handling such numerical integrations
across the sonic critical line, and try as much as we can to avoid
those unstable backward integrations from a nodal point and use up to
second-order derivatives\footnote{To avoid the dense singular points
in higher-order derivatives (see Hunter 1986 and our Appendix A), we
utilize only up to second-order derivatives and choose an appropriate
step away from a nodal point to assign initial values (see Appendix A).}
of Taylor expansions to estimate compatible initial values of $v(x)$
and $\alpha(x)$.

We first describe in some details the type 1 solutions. With the
Taylor expansion (21) for type 1 solutions near the sonic critical
line in the first quadrant, one obtains numerically a class of
solutions of $v(x)$ shown in Fig. 1 denoted by dashed curves. These
solutions of $v(x)$ in the first quadrant, referred to as {\it
plus} solutions by Shu (1977), approach negative constant values
of $V$ as $x\rightarrow+\infty$. The corresponding solutions of
$\alpha(x)$ scale as $A/x^2$ with $A<2$ as $x\rightarrow+\infty$,
consistent with the asymptotic behaviour (16) for $x\rightarrow
+\infty$. Following the same procedure, we also obtain type 1
solutions which pass through the sonic critical line in the fourth
quadrant. These solutions have the properties that $v(x)$ approaches
a positive constant $V$ and $\alpha(x)$ has the asymptotic behaviours
of $A/x^2$ with $A>2$ as $x\rightarrow+\infty$. Several examples are
shown in Table 2.
We will see that a class of discrete yet infinitely many type 1
solutions satisfies requirement (18) near the origin $x=0$. This
type of solutions were first derived by Hunter (1977). Note that
for $x_*=1$, the type 1 solution reduces to the static solution
(13) with $v(x)=0$.

We next discuss the type 2 solutions across the sonic critical
line. By asymptotic expansion (22) near the sonic critical point
in the first quadrant, we obtain another class of self-similar
solutions. 
Some of these solutions can go across the sonic critical line
smoothly once again in the fourth quadrant (either analytically or
with weak discontinuities) and approach constant values at large
$x$. Qualitatively, such solutions bear similar features as those
solid-line solutions. For comparison, we present the type 2
solutions in dotted curves in Fig. 1. Those type 2 solutions that
cannot pass twice the sonic critical line are regarded as
unphysical. Those type 2 similarity solutions that go across the
sonic critical line twice in the first and fourth quadrants are
new similarity solutions. There is yet another possibility of
matching shocks across the sonic critical line in the fourth
quadrant with different asymptotic solutions (e.g. Tsai \& Hsu
1995; Shu et al. 2002). We shall discuss continuous self-similar
solutions in Section 3.4.
\begin{figure}
\begin{center}
\includegraphics[scale=0.65]{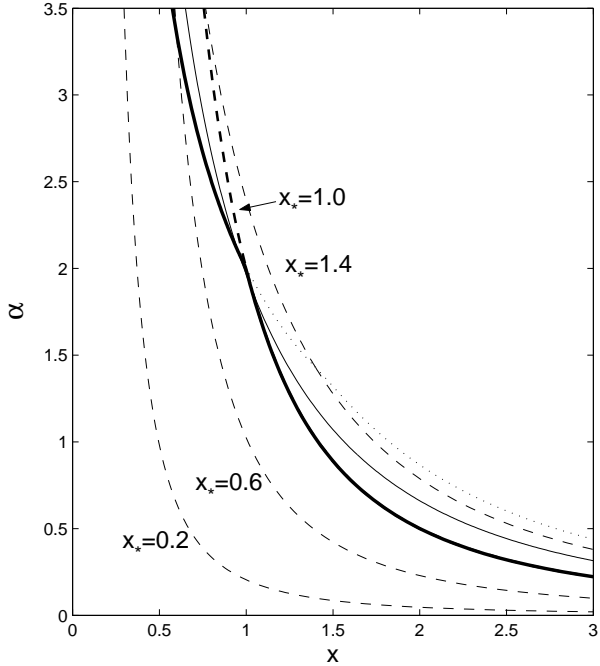}
\caption{For the same parameters given in Fig. 1, we show here the
corresponding solutions of reduced density $\alpha(x)$ by solid,
dotted and dashed curves, respectively. The light solid curve is
for $V=0$ and $A=3$ without crossing the sonic critical line. The
heavy solid line corresponds to the limiting case of $V=0$ and
$A=2^+$ (i.e. expansion-wave collapse solution). The dotted curve
is for one of the type 2 solutions that passes smoothly through the
sonic critical line twice (with weak discontinuities in the fourth
quadrant), one at $x_*=0.6$ and the other at $x_*=1.395$. The dashed
curves are type 1 solutions which pass through the sonic critical
line in the first and fourth quadrants, respectively. The heavy
dashed line is the type 1 solution with $x_*=1$ corresponding to
analytical static solution (13) of $\alpha=2/x^2$. Specifically,
the $x<1$ portion of the heavy solid line is the limit of the dotted
lines in the first quadrant as $x_*$ approaches $1^{-}$, while the
$x>1$ portion of the heavy solid line is approximately the static
solution almost completely overlaping with the heavy dashed curve.}
\end{center}
\end{figure}

\subsubsection{The Larson-Penston (LP) type and
the Hunter type of discrete solutions}

By both numerical and analytical analyses, Larson (1969a) and
Penston (1969a) independently derived a kind of monotonic
similarity solution, which has the asymptotic behaviour (18) near
the origin $x=0$. The LP-type of solution can be obtained
numerically by trying different values of $B$ parameter in
asymptotic solution (18) near $x=0$ and by integrating nonlinear
ODEs (9) and (10) outward from around the origin $x=0$ towards the
sonic critical line $x-v=1$ until critical condition (15) is met
such that the resulting solution matches with one of the two
eigensolutions at the sonic point. Here, $B$ is essentially an
eigenvalue parameter for nonlinear ODEs (9) and (10). For the
LP-solution, the eigenvalue $B$ is $\sim 1.67$. In earlier studies
(Shu 1977; Hunter 1977; WS), it was found that while the LP
solution is formally classified as type 2 solution (22) across the
sonic critical point in the fourth quadrant, the corresponding
eigensolution is in fact along the secondary direction for
$x_*>2$. That is, the LP-solution is analytic across the sonic
critical line. If we relax the requirement that solutions must be
analytic (as defined by Hunter 1986) across the sonic critical
line, it is then possible to construct a continuum band of
solutions with $0<B<1.67$ crossing the sonic critical line
involving weak discontinuities, with the LP-solution as the
limiting solution (i.e. the band 0 solutions constructed by WS
with $x_*\ge 2.33$, that have no stagnation points for $0<x<x_*$).
When $x_*>2$, the first-order derivatives of $v(x)$ for the two
eigensolutions are different with $D_2<D_1$, where $D$ stands for
the gradient of $v(x)$ at the sonic critical point and subscript
numerals of $D$ mark the types of eigensolutions. The LP-solution
has a speed gradient equal to $D_2$ as it intersects the sonic
critical line. Those solutions with $B\gsim 1.67$ have a speed
gradient smaller than $D_2$ and thus cannot go across the sonic
critical point because the required speed gradient falls within
the range of $[D_2,D_1]$ (WS).

We present several such examples in Fig. 5 with both $-v(x)$ and
$\alpha(x)$ shown simultaneously in the same plot [viz. $\alpha(x)$
in the upper portion and $-v(x)$ in the lower portion]. According
to the solution construction scheme of WS, there should exist an
infinite number of solutions fanning out from a single $x_*$ along
the primary direction across the sonic critical line that involve
weak discontinuities (e.g. second-order weak shocks as noted by Boily
\& Lynden-Bell 1995); for each case of Fig. 5, we plot only one
solution curve to the right of the sonic critical line for
illustration. Besides the LP-solution with $x_*=2.33$, the constant
$\alpha$ solution (14) with $x_*=3$ is also analytic at the sonic
critical point.
The former is along the secondary direction while the latter
is along the primary direction.
%

In the perspective of `complete solutions' ranging from
$t\rightarrow -\infty$ to $t\rightarrow +\infty$ to the collapse
problem, Hunter (1977) found, in addition to the LP-solution, a
class of infinitely many discrete similarity solutions (see his
fig. 2) that are analytic across the sonic critical line, labelled
by indices `a' and `c', `b' and `d' and so forth, although he
regarded the `a' and `c' solutions as physically invalid in the
scheme of `complete solutions'. These solutions satisfy asymptotic
condition (18) at the origin $x=0$ and are matched with type 1
eigensolutions at the sonic critical line with $x_*<2$ (along
secondary directions).
In our `semi-complete' perspective, each of Hunter type discrete
solutions consists of two separate branches in the first and
fourth quadrants as $x\rightarrow+\infty$. The first branch,
corresponding to the pre-catastrophic period (Hunter 1977) but
with opposite sign of $v(x)$, is demonstrated in Fig. 6; this
branch is obtained by numerically integrating from the origin
$x=0$ with $B$ parameter specified according to
expressions $(11a)-(11d)$
of Hunter (1977) [i.e. $B=\exp(Q_0)$ where $Q_0$ is a parameter
introduced by Hunter] to the sonic critical line and by further
expanding from the right side of the sonic critical line using
type 1 solution (20) along the secondary direction, until
$v(x)$ approaches a constant $V$ and $\alpha\rightarrow A/x^2$
as $x\rightarrow+\infty$. The second branch shown in Fig. 6,
according to the post-catastrophic period (Hunter 1977), is
obtained by specifying parameters $A$ and $-V$ in asymptotic
solution (16) and by integrating from $x\rightarrow+\infty$
back towards the origin $x=0$. The main difference is that we
focus on the first and fourth quadrants with positive $x$, while
Hunter (1977) studied such solutions with both negative and
positive time $t$. Hence, one complete solution such as `b' in
Hunter (1977), corresponds to two independent solution branches
in our semi-complete solution space. For example, the first
branch corresponding to the `b' solution of Hunter (1977) has
$|v|\rightarrow 0.295$ and $\alpha x^2\rightarrow 2.378$ as
$x\rightarrow+\infty$. By numerical integration from
$x\rightarrow+\infty$ under $V=-0.295$ and $A=2.387$, we obtain
the second branch labelled by `$b'$' shown in Fig. 6 also in
a heavy solid curve.

We note that Whitworth \& Summers (1985) have significantly
expanded Hunter's discrete similarity solutions to separate
continuous solution bands by allowing weak discontinuities at a
node along the primary direction or by local linear combination of
the two eigensolutions. The solution construction scheme of WS was
criticized by Hunter (1986) as being non-analytic across the sonic
critical line, even though such solutions are mathematically
possible. Similarity solutions with weak discontinuities across
the sonic critical line tend to be unstable (Hunter 1986; Ori \&
Piran 1988) and might bear little physical significance. In the
following analysis, we mainly focus on similarity solutions that
go across the sonic critical line analytically, although we mention
possible similarity solutions with weak discontinuities across the
sonic line for mathematical completeness.

\begin{figure}
\begin{center}
\includegraphics[scale=0.65]{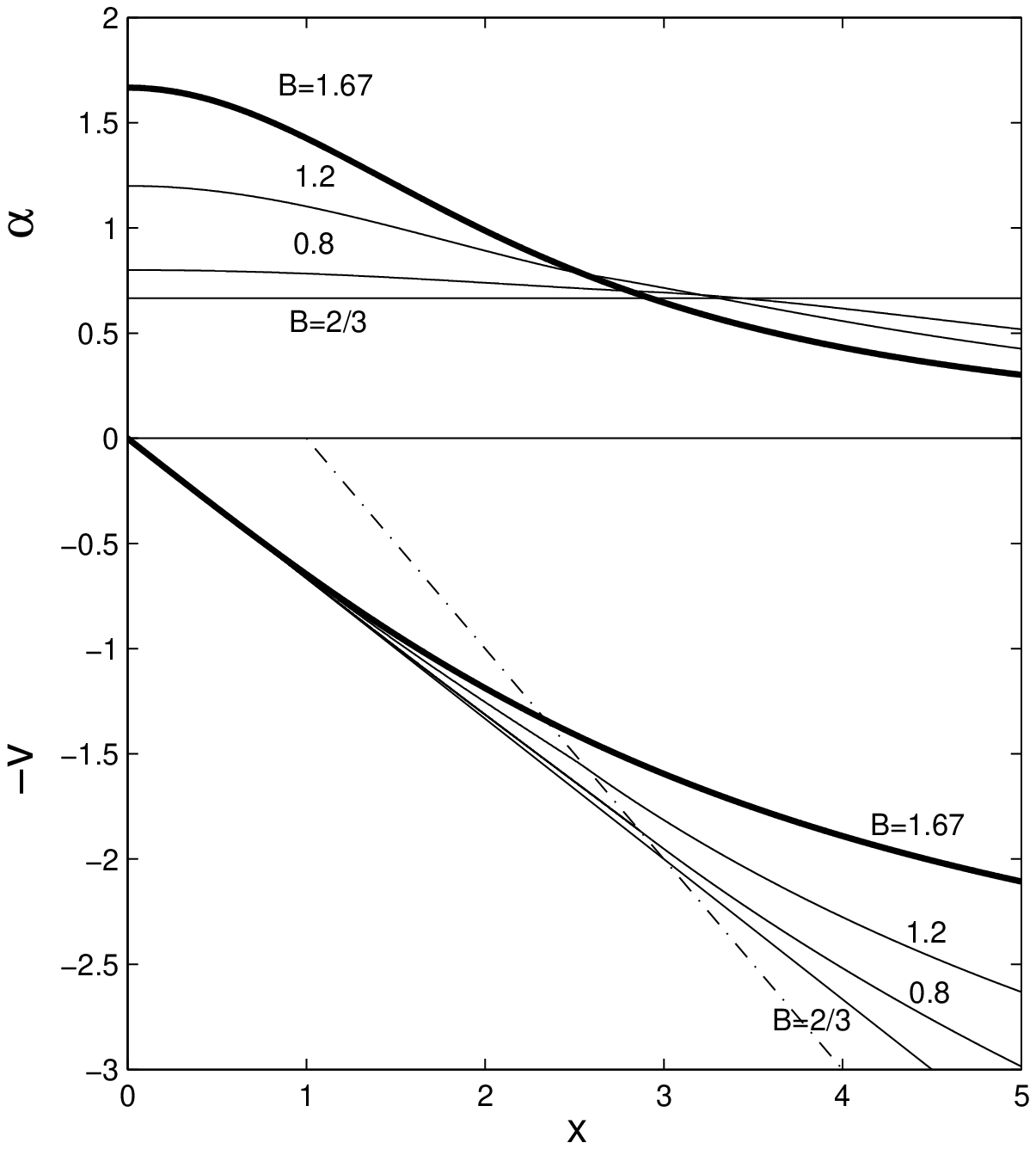}
\caption{The LP-type solutions. The solid lines beneath the $x$
axis are solutions of $-v(x)$ versus $x$, while those curves above
the $x$ axis are those of corresponding $\alpha(x)$ versus $x$.
The sonic critical line $x-v=1$ is denoted by the dash-dotted line
below the x-axis. The pair of heavy solid curves is the
LP-solution first found by Larson and Penston independently. These
LP-type solutions are obtained by integrating from the origin
$x=0$ with asymptotic solution (18) and assigning different values
for parameter $B$. Those values of $B$ in the range ($0, 1.67$]
can all satisfy critical condition (15) (corresponding to solution
band 0 with weak discontinuities of WS). The case of $B=2/3$
corresponds to the exact analytical solution (14). As
$x\rightarrow+\infty$, LP-type solutions all have the asymptotic
behaviours of (16) with different parameter combinations of $V$ and
$A$. }
\end{center}
\end{figure}

\begin{table*}{ }
\begin{tabular}{cccccccc}
\hline \hline $V$=0  & $A$ &2.0+ &2.2 &2.4 &2.6 &2.8 &3.0 \\
                     &$m_0$&0.976 &1.45 &1.885&2.311&2.743&3.186\\
\hline        $V$=-2 & $A$ &0.435 &0.5  &1.0    &1.5  &2.0    &2.5\\
                     &$m_0$&0.624 &0.804&2.117&3.594&5.219&6.971 \\
\hline
              $V$=2  & $A$ &6.0 &7.0&8.0&9.0&10.0&11.0\\
                     &
                     $m_0$&1.508&3.396&5.458&7.737&10.231&12.935\\
\hline
\end{tabular}
\caption{Parameter sets of $V$, $A$ and $m_0$ for self-similar
solutions without crossing the sonic critical line (see Figs. 2
and 3).}
\end{table*}

\begin{table*}{ }
\begin{tabular}{ccccccccc}
\hline\hline   $x_*$& 0.2 & 0.4 &0.6  &0.8  & 1 &1.2 &1.4 &1.6\\
\hline         $V$  &-2.82&-1.93&-1.22&-0.59&0  &0.57&1.12&1.67\\
               $A$  &0.177&0.468&0.864&1.374&2  &2.781&3.695&4.773\\
\hline
\end{tabular}
\caption{Key parameters $x_*$, $V$ and $A$ for the plus (or type
1) solutions in either first or fourth quadrant, where $x_*$ is
the location of the sonic critical point for the relevant
self-similar solutions. }
\end{table*}

\begin{figure}
\begin{center}
\includegraphics[scale=0.65]{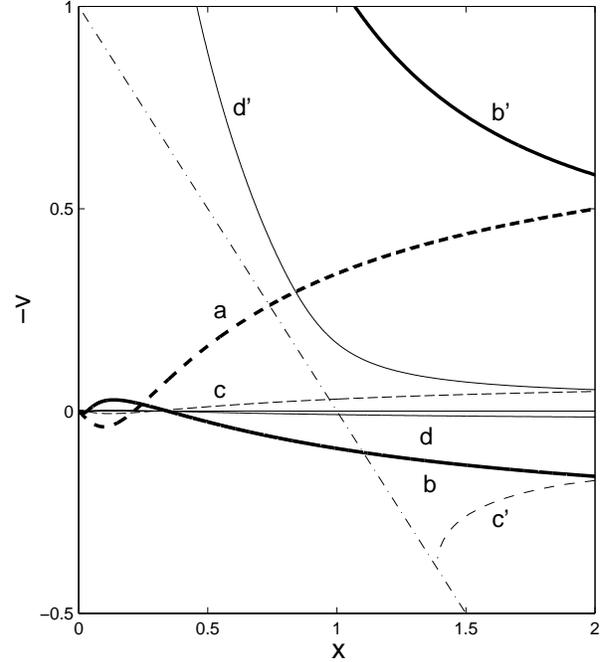}
\caption{Each of the Hunter-type discrete `complete solutions'
can be broken into two branches in the first and fourth quadrants
as $x\rightarrow+\infty$ in the `semi-complete' scheme. Such
similarity solutions are those of type 1 that satisfy condition
(18) with parameter $B$ specified by expressions (11a)-(11d) of
Hunter (1977), where $B\equiv\exp(Q_0)$ and $Q_0$ is a parameter
introduced by Hunter. Two such `complete solutions',
that is, the `b' and `d' solutions of Hunter (1977), appear
as two separate branches in heavy and light solid curves,
respectively. The corresponding `a' and `c' solutions of
Hunter (1977) shown in heavy and light dashed lines have only
one branch as the other branch cannot go across the sonic
critical line when integrating back from $x\rightarrow+\infty$.
For example, the branch marked `$c^\prime$' cannot pass the
sonic critical line smoothly. Further below the `$c^\prime$'
curve, to the lower right of the sonic critical line, would be
the `$a^\prime$' curve which is not shown here. The dash-dotted
line is the sonic critical line $x-v=1$.}
\end{center}
\end{figure}

\subsection{Solutions crossing the Sonic Line Twice}

We mentioned in subsection 3.3.2 that some of
the type 2 solutions with $x_*(1)<1$ in the first quadrant
can extend to the fourth quadrant and satisfy again critical
condition (15) when crossing the sonic critical line with
$x_*>1$. We denote the first critical point by $x_*(1)$ (less
than 1) in the first quadrant and the second critical point
by $x_*(2)$ (greater than 1). Such solutions, not necessarily
analytic at the second critical point $x_*(2)$, can be continued
across the sonic critical line towards $x\rightarrow+\infty$ in
an infinite group that consists of one solution along the
secondary direction and many other solutions along the primary
direction. In general, this kind of solutions along the primary
direction involve weak discontinuities across the sonic critical
line. Meanwhile, there exist solutions that match with one of
the two eigensolutions at $x_*(2)$ and that are thus analytic
across the sonic critical line. It is possible to construct
numerical solutions that are of the type 2 eigensolutions expanded
at $x_*(1)$ in the first quadrant and that are matched with either
type 1 or type 2 eigensolutions expanded at $x_*(2)$ in the
fourth quadrant.

\subsubsection{Solutions with weak discontinuities}

We now construct those similarity solutions which can go
across the sonic critical line once in the first quadrant
[$x_*(1)<1$] and again in the fourth quadrant [$x_*(2)>1$],
respectively. Apparently, such solutions must be of type 2
eigensolutions at the first critical point $x_*(1)$. There
are more possibilities at $x_*(2)$ for solution matchings.

As $x_*(1)$ in the first quadrant is a saddle point, numerical
integrations away from $x_*(1)$ are stable. We therefore integrate
from $x_*(1)$, using Taylor expansion (22) to select initial
values, towards the sonic critical line in the fourth quadrant.
When critical condition (15) is met as a solution curve of $v(x)$
versus $x$ crashes towards the sonic critical line $x-v=1$, the
solution can be extended from $x_*(2)$ further towards right with
increasing $x$. Many of such solutions have been found to satisfy
critical condition (15), although they may not be analytic at
different $x_*(2)$ (WS; Hunter 1986).

In particular, an infinite group of solutions will fan out from
$x_*(2)$ along the primary direction by the construction scheme of
WS. Such solutions do in fact involve weak discontinuities, and
are regarded by Hunter (1986) as unphysical\footnote{The physical
solutions are also required to be analytic at the critical point,
that is, only the two types of eigensolutions are analytic (see
Hunter 1986).}. Through our numerical explorations, there appear
to exist a continuum of such solutions in the approximate range
$0.2<x_*(1)<1$. When $x_*(1)$ becomes sufficiently small, the
solution of $-v(x)$ will turn upward as shown in Fig. 1 [e.g. the
heavy dotted-line with $x_*(1)=0.02$]. Solution behaviours of
$v(x)$ when $x_*(1)$ approaches the origin $x=0$ will be further
discussed later. Therefore, in a certain range of $x_*(1)$,
integrations from $x_*(1)$ can reach the nodal region $1<x<2$
along the sonic critical line where backward numerical
integrations from $x_*(2)$ are unstable (i.e. extremely sensitive
to the accuracy of starting conditions) for type 2 solutions
(along the primary direction) and are neutrally stable for type 1
solutions (along the secondary direction).

\begin{figure}
\begin{center}
\includegraphics[scale=0.45]{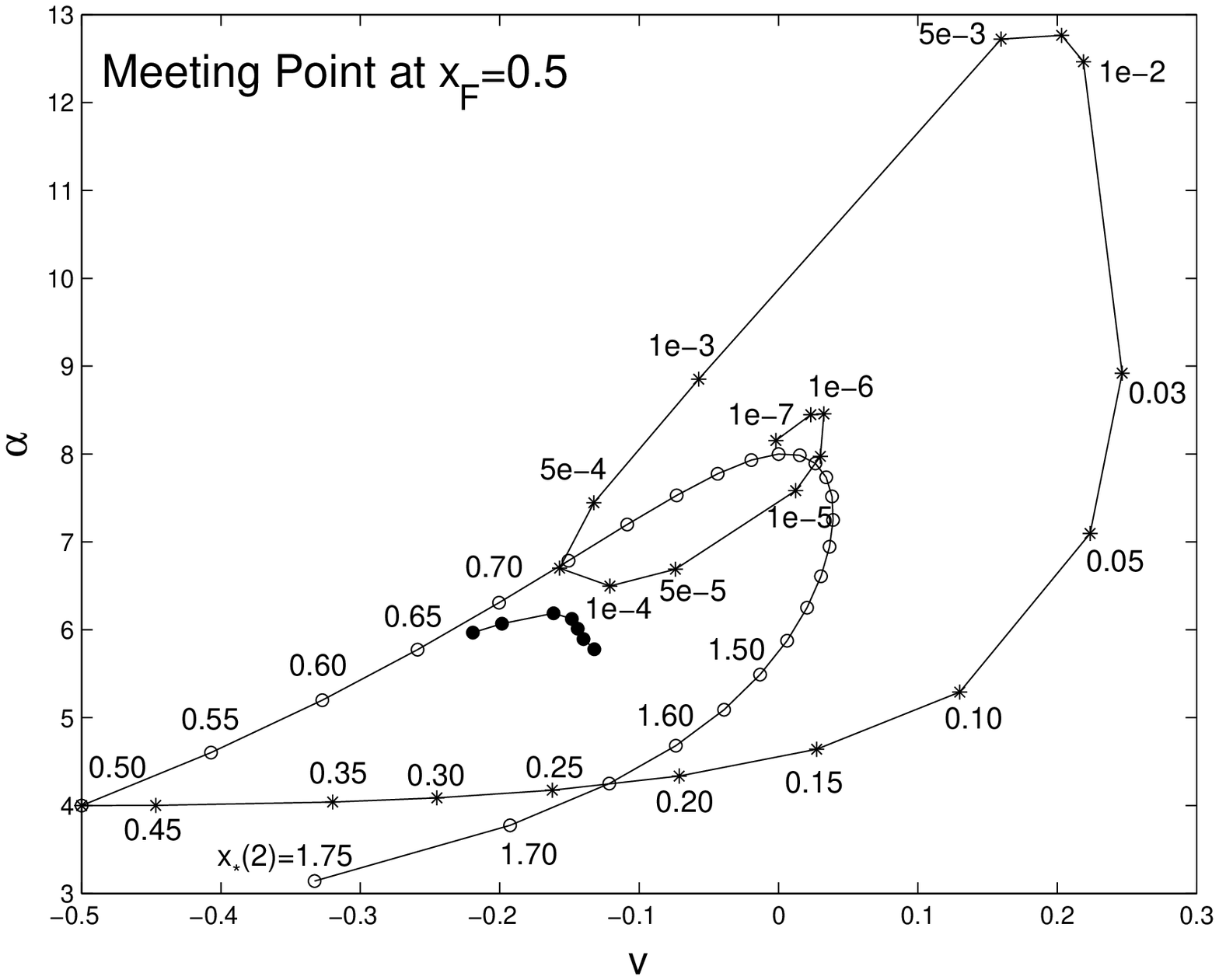}
\caption{The phase diagram of $v$ versus $\alpha$ at a chosen
meeting point $x_F=0.5$ for type 2-type 1 solution matches [that
is, type 2 eigensolution at $x_*(1)$ and type 1 eigensolution at
$x_*(2)$ along the secondary direction for $x_*(2)<2$]. Each
asterisk symbol $*$ denotes an integration from $x_*(1)$ of a type
2 eigensolution in the first quadrant and each symbol circle
$\circ$ denotes an integration from $x_*(2)$ of a type 1
eigensolution in the fourth quadrant. The $*$ curve and the
$\circ$ curve intersect with each other in the phase diagram. From
the spiral trend of the $*$ curve (somewhat similar to fig. 2 of
Hunter 1977), we infer that there exist an infinite number of
discrete type 2-type 1 self-similar solutions with the static
solution (diverge at the origin) as the limiting case. The first
such solution match corresponds to $x_*(1)\approx 0.23$ and
$x_*(2)\approx 1.65$. We have chosen the meeting point $x_F$
to be 0.5 here. For $x_{*}(1)=x_{*}(2)=0.5$ with the same set of
$\alpha$ and $v$ in the phase diagram, there are two different
sets of derivatives given by expressions (19) and (20). Despite
the appearance, the phase point (-0.5, 4) for $(v,\alpha)$ at
$x_{*}(1)=x_{*}(2)=0.5$ in this figure is therefore not a real
match because the first-order derivatives are distinctly different.
We add the locus of the type 2 backward integrations from
$x_*(2)$ in solid circles.}
\end{center}
\end{figure}

\begin{figure}
\begin{center}
\includegraphics[scale=0.45]{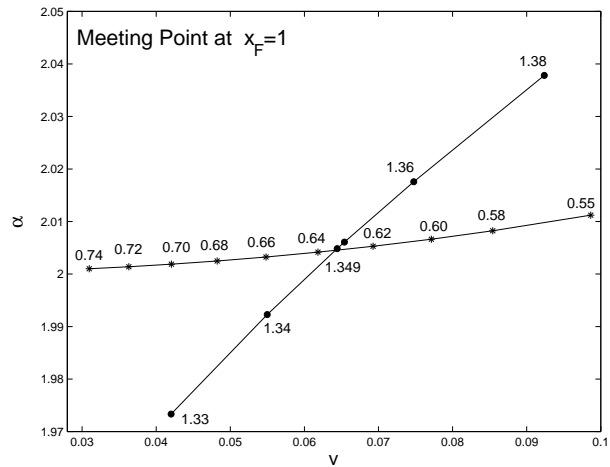}
\caption{The phase diagram of $v$ versus $\alpha$ at a chosen
meeting point $x_F=1$. Each asterisk symbol $*$ denotes an
integration from $x_*(1)$ with its value shown and each full
circle denotes an integration from $x_*(2)$ with its value
shown. Both integrations are of type 2 eigensolutions when
crossing the sonic critical line at $x-v=1$. There is an
intersection at $x_F=1$ in the phase diagram and thus a unique
match of $v(x_F)$ and $\alpha(x_F)$ from both sides of $x_F=1$.
For this matched solution, we have $x_*(1)\approx 0.632$ and
$x_*(2)\approx 1.349$. In searching for such matchings in the
phase diagram, there exist certain values of $x_*(2)$ starting
from which numerical solutions $v(x)$ and $\alpha(x)$ cannot
reach $x_F=1$; the corresponding $v(x)$ will curve upward and
quickly run into the sonic critical line again.}
\end{center}
\end{figure}

\begin{figure}
\begin{center}
\includegraphics[scale=0.45]{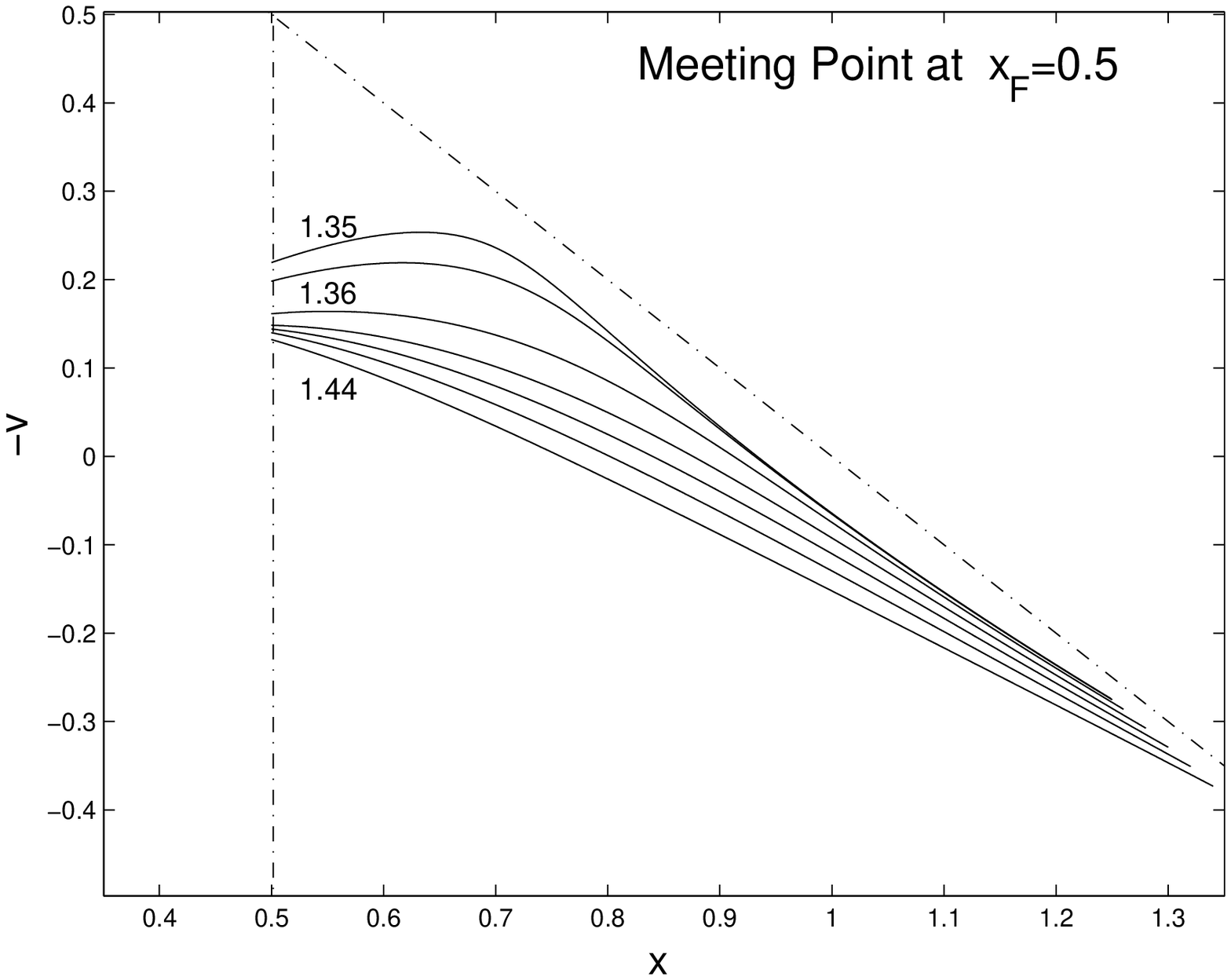}
\caption{Backward integrations from a type 2 eigensolution at
$x_*(2)$; those that can reach $x_F=0.5$ are shown. The labelled
numbers indicate the values of $x_*(2)$. The corresponding locus
of $\alpha$ versus $v$ are shown in Fig. 7 as solid circles.}
\end{center}
\end{figure}


\subsubsection{Self-similar solutions that
are analytic across the sonic critical line}

Described below is a useful numerical procedure of finding a
similarity solution that is analytic across the sonic critical line
(Hunter 1977). By properly specifying pertinent parameters, one
integrates from a series of starting points near and alongside a
portion of the sonic critical line in the first quadrant towards a
chosen meeting point $x=x_F$, plots the resulting series values of
functions $v$ and $\alpha$ in pair at $x=x_F$ in the $v$-versus-$\alpha$
phase diagram to obtain the first locus. One then starts integrations
near and alongside another portion of the sonic critical line in the
fourth quadrant by specifying eigensolutions as initial values towards
the same $x_F$ and thus obtains the second locus in the $v-\alpha$
phase diagram. When the two phase loci intersect with each other, one
then finds a self-similar solution that goes across the sonic critical
line twice analytically. Note such an intersecting point in the phase
diagram should be at a $x-$location away from the sonic critical line.
The derivatives of $v(x)$ and $\alpha(x)$ are uniquely determined by
the values of $v$ and $\alpha$ by equations (9) and (10) for an
integration point $x$ away the sonic critical line; along the sonic
critical line, there are two distinct pairs of derivatives of $v(x)$
and $\alpha(x)$ given by equations (19) and (20).

In these numerical integrations, we essentially follow the same
method of Hunter (1977). The main difference is that the assigned
points of Hunter are now replaced by different values of $x_*(1)$
for type 2 eigensolutions at saddle points along the sonic
critical line in the first quadrant. Numerical integrations away
from a saddle point $x_*(1)$ are stable and can be performed in a
straightforward manner. In contrast, numerical integrations
backward from a nodal point $x_*(2)$ near and alongside the sonic
critical line in the fourth quadrant require extra care. In
practice, we use either type 1 or type 2 eigensolutions around
$x_*(2)$ up to the second-order derivatives and a carefully chosen
step backward
of the sonic critical line to assign initial values. We then integrate
towards $x_F$ from both $x_*(1)$ and $x_*(2)$ to search for solution
matches described above. Note that the relevant values of $x_*(2)$
are less than 2 as no integrations from $x_*(1)$ can reach the range
of the sonic critical line with $x>2$.

For a type 2-type 1 solution match, we choose\footnote{Once the
meeting point $x_F$ is chosen, only certain ranges of $x_*(1)$ and
$x_*(2)$ are pertinent to a specific type of similarity solutions
being searched for. Typically, values of $x_*(1)$ lie in the
interval of $(0, x_F)$ and values of $x_*(2)$ cannot be too large.
Otherwise, integrations away from $x_*(2)$ may not reach $x_F$ and
can become extremely sensitive to initial values along the primary
direction at the sonic point. Numerical integrations from
$x_*(1)\geq 0.5$ with type 2 eigensolutions in the first quadrant
are found to approach $x_*(2)$ in the type 2 direction (i.e. the
primary direction) and it is thus not easy to match with the type
1 eigensolutions along the sonic critical line in the fourth
quadrant.} $x_F=0.5$ and integrate from $x_*(1)$ with a type 2
eigensolution. Meanwhile, we integrate from $x_*(2)$ with a type 1
eigensolution (i.e. along the secondary direction) towards $x_F$.
The $v-\alpha$ phase diagram at $x=x_F$ is shown in Fig. 7. The
`spiral structure' is strikingly similar to that in fig. 2 of
Hunter (1977), the trend of which then implies an infinite number
of discrete type 2-type 1 self-similar solutions that can pass
twice the sonic critical line analytically. The first three of
such similarity solutions are displayed in the form of $-v(x)$
versus $x$ in Fig. 10, with the enlarged part shown in Fig. 11 for
small $x$ using a logarithmic scale. It is interesting to note
that the number of stagnation points, corresponding to
self-similar radial oscillations, increases for such class of
similarity solutions with smaller $x_*(1)$. Specifically, the
first similarity solution (labelled by 1) has only one stagnation
point crossing the $x$-axis and the relevant parameters are
$x_*(1)\approx 0.23$ and $x_*(2)\approx 1.65$. The parameters of
the second solution are $x_*(1)\approx 2.5858\times 10^{-4}$ and
$x_*(2)\approx 0.743$. The parameters of the third solution are
$x_*(1)\approx 6\times 10^{-6}$ and $x_*(2)\approx 1.1$.

\begin{figure}
\begin{center}
\includegraphics[scale=0.65]{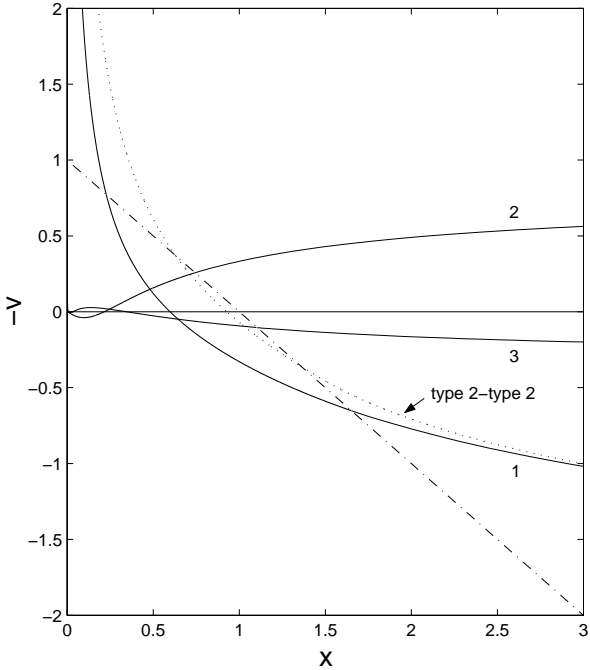}
\caption{The first three discrete type 2-type 1 self-similar
solutions [$v(x)$ versus $x$ all denoted by light solid curves]
crossing twice the sonic critical line, the dash-dotted line for
$x-v=1$, analytically and one type 2-type 2 similarity solution,
denoted by the dotted curve, that also crosses twice the sonic
critical line. Numerals $1,2,3$ labelled along the three solid
curves denote the number of stagnation points (i.e. $v=0$) at the
$x$-axis. All these similarity solutions diverge, approaching
minus infinity as $x\rightarrow 0^{+}$ that are displayed more
explicitly in Fig. 11 with the $x$-axis in a logarithmic scale.}

\end{center}
\end{figure}
\begin{figure}
\begin{center}
\includegraphics[scale=0.65]{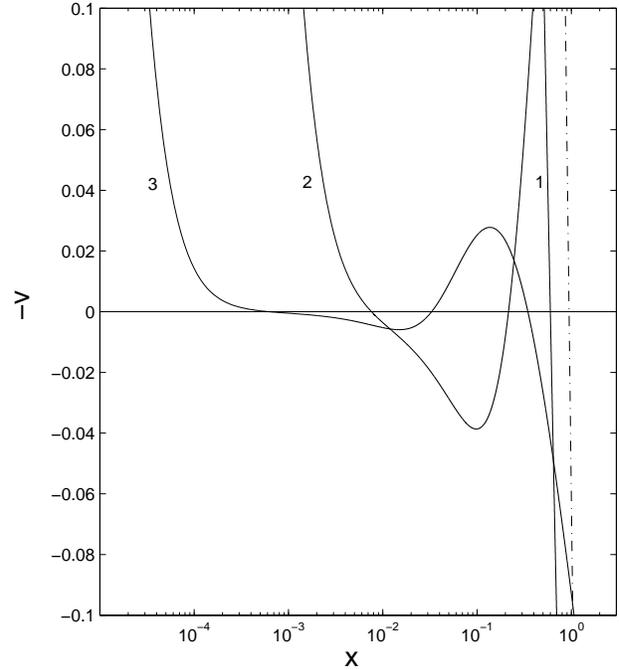}
\caption{The enlarged portions of solution curves near the origin
$x=0$ of Fig. 10, showing the diverging behaviours and the number
of stagnation points of discrete type 2-type 1 similarity
solutions as $x\rightarrow 0^{+}$. The $x$-axis is shown in a
logarithmic scale here. The dash-dotted line is the sonic critical
line $x-v=1$. Physically, these profiles represent subsonic radial
oscillations in a self-similar manner. }
\end{center}
\end{figure}

The situation becomes somewhat involved for a type 2-type 2
solution match as $x_*(2)$ in this case is a nodal point and the
type 2 eigensolutions with $x_*(2)<2$ happen to be along the
primary direction (Jordan \& Smith 1977). Only within a narrow
range of $x_*(2)$, can numerical solutions be integrated to reach
$x_F=0.5$. For this reason, we choose different values of $x_F$.
First, we choose $x_F=1$ and integrate from both $x_*(1)$ and
$x_*(2)$ by type 2 eigensolution (22). The corresponding
$v-\alpha$ phase diagram is displayed in Fig. 8. It is clear that
there exists a discrete solution match between $x_*(1)\approx
0.632$ and $x_*(2)\approx 1.349$. By selecting different meeting
point $x_F$, one can search for other ranges of $x_*(1)$ and
$x_*(2)$ for possible solution matches but none was found in our
numerical exploration. The $v-\alpha$ phase locus of those
numerical integrations that can reach $x_F=0.5$ come very close to
but do not intersect the $v-\alpha$ phase locus determined from
$x_*(1)$ (see Figs. 7 and 9). In short, the solution which passes
consecutively the first sonic critical point at $x_*(1)\approx
0.632$ in the first quadrant and the second sonic critical point
at $x_*(2)\approx 1.349$ in the fourth quadrant is a unique type
2-type 2 self-similar solution being analytic at both sonic
critical points; and this solution has only one stagnation point
across the $x$-axis. For completeness, we further integrate from
the relevant $x_*(2)$ towards $x\rightarrow +\infty$ in terms of
the type 2 solution (see dotted curve in Fig. 10).

By numerical exploration and analysis, we obtained discrete EECC
self-similar solutions within the entire range of $0<x<+\infty$,
each crossing twice the sonic critical line analytically in the
first and fourth quadrants. For all these self-similar solutions,
the $x\rightarrow 0^{+}$ profiles have asymptotic diverging
behaviours (17) and the $x\rightarrow +\infty$ profiles have
asymptotic behaviours (16). The basic solution profile of the
first (labelled by numeral 1 in Fig. 10) type 2-type 1 similarity
solution with one stagnation point is qualitatively similar to its
counterparts without crossing the sonic critical line as shown in
Fig. 3 (i.e. curves with $V=2$, $A=6, 8, 10$, respectively).
Furthermore, the family of type 2-type 1 solutions is very similar
to Hunter's solution family; two major differences are: (1) the
condition at the origin $x=0$ has been changed from $v\rightarrow
0$ to $v\rightarrow -\infty$; (2) the solutions are obtained in the
`semi-complete' space instead of `complete' space. In contrast to
the `accretion solutions' free of any sonic critical point described
in subsection 8.2 of Boily \& Lynden-Bell (1995), our novel EECC
similarity solutions go across the sonic critical line twice and
apparently have been missed by earlier investigations on isothermal
collapses and flows with spherical symmetry.

We note again that a numerical integration backward from $x_*(2)$
along the primary direction is extremely sensitive to initial
values. Thus, the accuracy of the type 2-type 2 similarity
solution obtained may be improved, and there might still be
matches at $x_F=0.5$ if other initial steps away from $x_*(2)$
were chosen. For example, if one chooses a step of $0.001$ away
from $x_*(2)$ and integrates backward using a type 2 eigensolution
to $x_F=0.5$, it might be possible to find one or more
intersections with those integrated from $x_*(1)$. Furthermore,
solutions along the primary direction, with so many possible
non-analytic ones tangential to the same direction, might be
unstable (e.g. Ori \& Piran 1988; Lai \& Goldreich 2000) and might
be physically irrelevant. In Section 4, we shall mainly focus on
the type 2-type 1 self-similar solutions with eigensolutions along
the secondary direction at $x_*(2)$.

\section{Envelope Expansion with Core Collapse (EECC) solutions}

We have examined the similarity solution structure in Section 3.
We now divide all solutions extended through $0<x<+\infty$
into two major classes. Among the two classes, each solution is
continuous and smooth within the entire range of $x$. For $x$
approaches infinity, a solution is determined by solution (16),
that is, the reduced velocity $v(x)$ approaches a constant value
$V$ and the reduced mass density $\alpha(x)$ scales as $A/x^2$
asymptotically; when $x$ approaches the origin, the solutions are
determined either by solution (17) as a free-fall state with a
reduced core mass $m_0$ or by condition (18) of vanishing velocity
and finite $\alpha$. The first group is referred to as Class I
similarity solution and the second group is referred to as Class
II similarity solution. For the structure of possible similarity
solutions, we focus on Class I similarity solutions.


The Class I similarity solutions can be further divided into three
subclasses based on the signs of the asymptotic radial speed
parameter $V$. We denote by Class Ia similarity solutions for
$V>0$, by Class Ib similarity solutions for $V=0$ and by Class Ic
similarity solutions for $V<0$, respectively. The Class Ib
similarity solutions were studied previously by Shu (1977) as
infall similarity solutions with an important limiting solution,
the `expansion-wave collapse' solution. Recently, these solutions
(with $V=0$ and $A<2$) have been used to construct shocked
LP-solutions to model `champagne flows' in H{\sevenrm II} regions
surrounding new-borne stars (Shu et al. 2002). The Class Ic
similarity solutions are core-collapse solutions with inflows at
large radii. In contrast, the existence of Class Ia similarity
solutions is a new revelation. Such similarity solutions describe
situations that for a given $r$, the radial flow may be outward at
a certain time, undergo one or more oscillations,
and eventually collapse towards the central core in a nearly
free-fall manner. It also describes a spherical fluid system
under self-gravitation at a given instant $t$ that its interior
is core collapsing while its exterior envelope is expanding with
or without subsonic radial oscillations in between. For this
reason, we refer to the Class Ia solutions as `envelope expansion
with core collapse' (EECC) solutions.\footnote{
In describing early models of galactic winds,
Holzer \& Axford (1970) discussed steady-state solutions that may
involve concurrent outflows and inflows given distributed sources of
mass, momentum and energy (e.g. Burke 1968; Johnson \& Axford 1971).
}


\subsection{Physical Interpretations}

We now focus on the physical meaning of the EECC self-similar
solutions and examine specifically such EECC solutions with only
one stagnation point along the $x-$axis. In the scenario of Shu
(1977), the `expansion-wave collapse' solution is constructed by
joining two solutions at $x_*=1$ with $\alpha(x)$ being continuous
there. The $x<1$ portion is the type 2 solution touching the sonic
critical line at $x_*=1$ and the $x\geq 1$ portion is simply the
hydrostatic solution for a singular isothermal sphere (SIS). Such
a solution physically describes a situation with a collapsing core
approaching a free-fall state in the central region and with a
hydrostatic exterior at rest for $x\geq 1$. The spherical surface
separating the interior and exterior expands outward in a
self-similar manner, at the speed of the isothermal sound $a$.

With this basic information in mind, the EECC self-similar
solutions can be understood as follows. The interior portion
($x<\xi$, where $\xi$ is the location of a surface of separation)
is the infall/collapse region while the exterior portion ($x\geq
\xi$) is the expansion region toward infinity. The outward
movement of the exterior envelope leads to a wind with a constant
speed at large radii. The spherical surface of separation $r_0=\xi
at$ where the gas remains at rest travels outward as time goes on;
this feature is similar to the `expansion-wave front' of Shu (1977).
The `expansion-wave collapse' solution of Shu may also be regarded
as a limiting case of our new EECC similarity solutions in the sense
that the outer envelope is static at large $x$ asymptotically rather
than expanding or contracting at constant speeds. The travel speed
of the expanding surface of separation is $\xi a$, which is either
subsonic for $\xi<1$ [in cases of those EECC similarity solutions
(e.g. Figs. 10 and 11) crossing the sonic critical line twice] or
supersonic for $\xi >1$ [in cases of those EECC similarity solutions
(e.g. Fig. 3) that do not intersect the sonic critical line].

This explanation can be further extended to those EECC similarity
solutions with more than one stagnation points, involving radial
oscillations in the subsonic region. For such EECC similarity
solutions, there exist more than one spherical surfaces of
separation that expand outward with time. The distances between
spherical surfaces of separation increase with time in a
self-similar manner.

\subsection{Astrophysical Applications}


All similarity solutions in the range of $0^{+}<x< +\infty$ are
derived from nonlinear hydrodynamic equations and can be applied to
various astrophysical systems with proper contexts. With necessary
qualifications and adaptations, such EECC similarity solutions,
which describe an outgoing envelope and a collapsing core may be
applied to the late evolution stage of stars, for example, the
asymptotic giant branch (AGB) phase or post-AGB phase before the
appearance of a planetary nebula system. We have the following
physical scenario in mind. A star (with a progenitor less than a
few solar masses) swells enormously and sustains a massive wind
with an asymptotic speed of $\sim 10-20\hbox{ km s}^{-1}$. With
an insufficient nuclear fuel supply from a certain epoch on, the
central region starts to collapse while the outer envelope
continues to expand. In a concurrent manner, the outer expansion
removes stellar envelope mass, while the central infall and collapse
produce a proto white dwarf at the center. Sufficiently away from
initial and boundary conditions, the system may gradually evolve
into a dynamic phase describable by an EECC similarity solution
during a timescale of a few hundred to several thousand years.
Within this rough range of timescales, the accumulation of central
core mass should not exceed the Chandrasekhar mass limit of
1.4$M_{\odot}$ for a remnant white dwarf to survive; otherwise, a
central explosion may occur within a planetary nebula. Meanwhile,
the outer envelope expansion during the EECC phase can be faster
than the pre-existing massive wind. This process will generally
lead to formation of outward moving shock when a faster EECC outflow
catches up a slower pre-existing wind. Theoretically, it is possible
to construct shocked EECC similarity solutions to generalize the prior
model results of Tsai \& Hsu (1995) and Shu et al. (2002). We therefore
suggest that the dynamical evolution of an EECC phase of around or less
than a few thousand years may be the missing linkage between the AGB or
post-AGB phase and the gradual appearance of a planetary nebula.
Depending on physical parameters of the progenitor star, it might
happen that the core collapse and subsequent central infalls during
an EECC phase lead to a core mass exceeding the Chandrasekhar
$1.4M_{\odot}$ and thus induce an explosion with an intensity
determined by the actual rate of mass infall.

We now explore the utility of these EECC solutions obtained in this
paper, including those that pass through the sonic critical line twice
(Figs. 10 and 11) and those that do not involve critical points (Fig. 3).
Relevant sets of parameters $\{A,  V,  m_0\}$ are summarized in Table 3
and the corresponding similarity solution curves of $\alpha(x)$, $v(x)$
and $m(x)$ versus $x$ are presented in Figs. 12-15 (those curves for
NoCP2, very close to the ones for NoCP3, were omitted to avoid
cluttering). Those similarity solutions of `expansion wave' type
without involving critical points appear to give reduced core masses
greater than $0.976$ in general. In comparison, the type 2-type 1 EECC
solutions can give rise to reduced core masses less than $0.976$. For
the type 2-type 1 EECC similarity solutions with more than one
stagnation points (see Fig. 11), the reduced core mass parameter
$m_0$ can be very small.

\begin{figure}
\begin{center}
\includegraphics[scale=0.65]{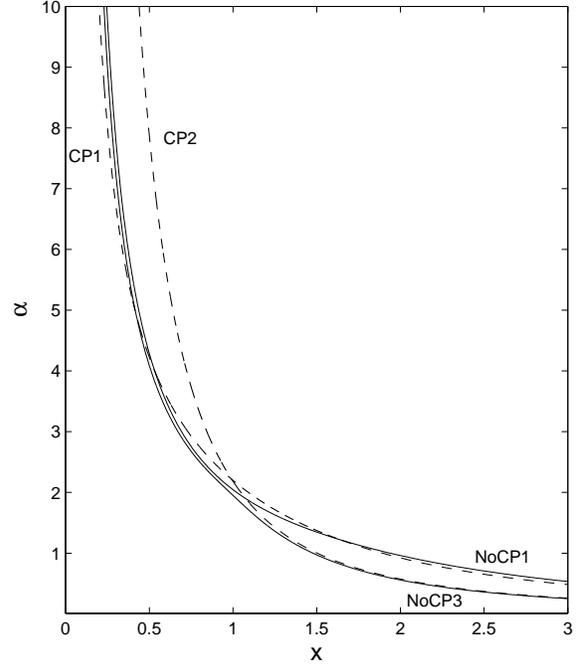}
\caption{The curves of reduced density $\alpha$ versus $x$ for
several EECC similarity solutions in linear scales. The light
solid curves NoCP1 and NoCP3 stand for those without crossing a
sonic critical point, and the dashed curves CP1 and CP2 denote
those crossing the sonic critical line twice, that is, the first
and the third type 2-type 1 similarity solutions of Fig. 10. Note
that the two light solid curves NoCP1 and NoCP3 are very close to
each other for $x<1$.}
\end{center}
\end{figure}
\begin{figure}
\begin{center}
\includegraphics[scale=0.65]{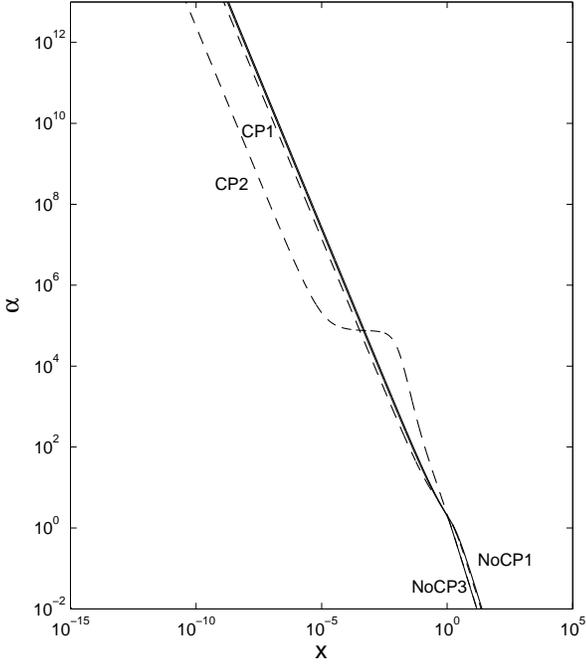}
\caption{The curves of reduced density $\alpha$ versus $x$ for the
same EECC similarity solutions in Fig. 12 but in logarithmic
scales. Note that $\alpha\rightarrow 0^{+}$ for $x\rightarrow
+\infty$ and $\alpha\rightarrow +\infty$ for $x\rightarrow 0^{+}$.
The rates of the asymptotic behaviours are different. The apparent
wiggle in $\alpha(x)$ of CP2 dashed curve corresponds roughly to
the range of the three stagnation points of type 2-type 1 solution
labelled by numeral 3 in Figs. 10 and 11.} 
\end{center}
\end{figure}

\begin{figure}
\begin{center}
\includegraphics[scale=0.65]{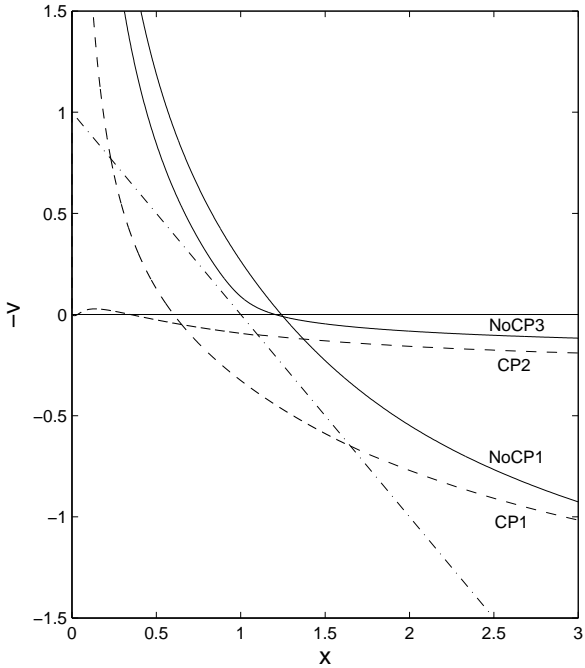}
\caption{The curves of $-v$ versus $x$ for the same EECC similarity
solutions of Fig. 12 in linear scales. In addition to the two
type 2-type 1 similarity solutions labelled by 1 and 3 of Fig. 10
displayed in dashed curves here, we also show two solutions without
crossing the sonic critical point in light solid curves. The curve
$v(x)$ for CP2 goes to $-\infty$ as $x\rightarrow 0^{+}$, as shown
in Fig. 11.}
\end{center}
\end{figure}

\begin{figure}
\begin{center}
\includegraphics[scale=0.65]{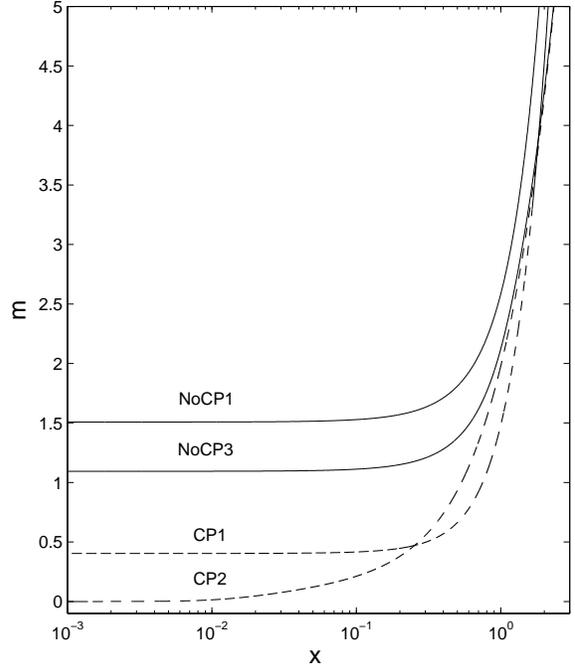}
\caption{The curves of reduced enclosed mass $m$ versus $x$ for
the same EECC similarity solutions of Fig. 12 in logarithmic
scale for $x$ axis. Each approaches a finite core mass parameter
$m_0$ as $x\rightarrow 0^{+}$ (see Table 3 for relevant parameters).}
\end{center}
\end{figure}

During the EECC similarity evolution, the core mass increases with
time until the termination of the self-similar phase by other
processes. To crudely estimate the final core mass near the end of
the EECC phase, we have $G=6.67\times 10^{-8} \hbox{ cm}^3\hbox{
g}^{-1}\hbox{ s}^{-2}$ and a typical timescale of
$t=3\times10^{10}\hbox{ s}\sim 10^{3}\hbox{ yr}$ for the period
between a late AGB phase and a beginning planetary nebula phase
(e.g. Balick \& Frank 2002). By transformation (5), the final core
mass is given by $M(0,t)=a^3tm_0/G$. For an isothermal sound speed
$a=20\hbox{ km s}^{-1}=2\times 10^{6}\hbox{ cm s}^{-1}$, the final
core mass is $M(0,t)\sim 2\times 10^{3}m_0\ M_{\odot}$. For $a$ of
the order of several $\hbox{km s}^{-1}$, the final core mass is
approximately $\sim m_0M_{\odot}$. Qualitatively, a smaller $m_0$
or a shorter period of the EECC phase would be necessary for a
higher sound speed $a$ without exceeding the Chandrasekhar mass
limit of $1.4M_{\odot}$ for the central proto white dwarf. Within
this scenario, it appears that type 2-type 1 EECC similarity
solutions with more subsonic stagnation points are more likely of
practical interest. For the CP2 EECC similarity solution (see Figs.
10, 11, 14 and Table 3), the outer envelope expands at large radii
at a speed of $\sim 0.3$ times the sound speed (i.e. several
$\hbox{ km s}^{-1}$); the corresponding reduced core mass
parameter $m_0$ is $1.2\times 10^{-5}$.

For the problem at hand, the energy density ${\cal E}$ and
the energy flux density ${\cal F}$ can be readily identified
from the energy conservation equation (4)
\begin{equation}
\begin{split}
&{\cal E}=\frac{\rho u^2}{2} +a^2\rho
\bigg[\ln\bigg(\frac{\rho}{\rho_c}\bigg)-1\bigg] -\frac{1}{8\pi G}
\bigg(\frac{\partial\Phi}{\partial r}\bigg)^2\ ,\\
&{\cal F}=u\rho\bigg[\frac{u^2}{2}
+a^2\ln\bigg(\frac{\rho}{\rho_c}\bigg)\bigg]\ .
\end{split}
\end{equation}
By using similarity transformation (5) for dependent
variables and relations (7) and (12), we further obtain
\begin{equation}
\begin{split}
&{\cal E}=\frac{a^2}{8\pi Gt^2}\bigg[\alpha
v^2+2\alpha\bigg(\ln\frac{\alpha}{4\pi
Gt^2\rho_c}-1\bigg)-\alpha^2(x-v)^2\bigg]\ ,\\
&{\cal F}=\frac{a^3}{8\pi Gt^2}
\bigg(\alpha v^3+2\alpha v\ln\frac{\alpha}{4\pi Gt^2\rho_c}\bigg)\ .\\
\end{split}
\end{equation}
For an actual astrophysical system, one may then use equations (23)
and (24) to estimate relevant energy density and energy flux density,
respectively.


\begin{table}
\begin{center}
\begin{tabular}{cccc}
\hline \hline Description  &$V$   & $A$   &$m_0$\\
               NoCP1       &2.000&6.000& 1.508\\
               NoCP2       &0.200&2.500& 1.548\\
               NoCP3       &0.200&2.300& 1.094\\
                CP1        &1.845&5.158& 0.406\\
                CP2        &0.287&2.371& $1.200\times 10^{-5}$\\
\hline \hline
\end{tabular}
\caption{Parameters $V$, $A$ and $m_0$ for the EECC similarity
solutions; CP for solutions involving critical points and NoCP for
solutions without critical points.}
\end{center}
\end{table}

\section{Discussion}

In this paper, we investigated the spherical self-similar
solutions for self-gravitating isothermal flows in the
`semi-complete solution space' extending from the initial instant
$x\rightarrow +\infty$ to the final instant $x\rightarrow 0^{+}$.
The relevant similarity solutions are obtained and classified to
compare with previous solutions and analyses [Larson 1969a;
Penston 1969a; Shu 1977; Hunter 1977; Whitworth \& Summers 1985
(WS)].

The novel class of infinitely many similarity solutions obtained
in this paper, i.e., the EECC self-similar solutions and the
envelope contraction with core collapse (ECCC) self-similar
solutions (e.g. solution labelled 2 in Figs. 10 and 11 that passes
the sonic critical line twice with specific parameters $V=-0.766$,
$A=1.209$ and $m_0=5.171\times 10^{-4}$), is constructed in the
`semi-complete solution space'. This is to be compared with the similarity
solutions derived in the `complete solution space' first introduced by
Hunter (1977). By the invariance property under time reversal in this
classical problem, these new self-similar solutions may be formally
extended to the `complete solution space' by properly joining two separate
branches in the first and fourth quadrants of the `semi-complete solution
space' but with diverging asymptotic behaviours of $v(x)$ and $\alpha(x)$
at $t\rightarrow -\infty$. Similarity solutions without crossing the sonic
critical line as shown in Figs. 2 and 3 also carry ECCC and EECC features,
respectively, and are much easier to construct technically in the isothermal
approximation; in a polytropic gas with various laws of radiative losses,
some qualitatively similar features were mentioned in subsection 8.2 of
Boily \& Lynden-Bell (1995).

Regarding the isothermal approximation, the core collapsing (CC)
portion may be justified qualitatively on the ground that the gas
would heat up by adiabatic compressions. In the presence of a
sufficient amount of radiative agents in a relatively high density
region involving more frequent collisions, such radiative cooling
processes may sustain an approximate condition of quasi-constant
temperature. To sustain an isothermal condition for the envelope
expansion (EE) portion, some heating mechanisms are necessary to
compensate the inevitable cooling resulting from adiabatic
expansions. Naturally, radiations from gas in the CC portion can
be absorbed by gas in the EE portion to heat up the latter to some
extent (depending on the escape efficiency of photons). In the
astrophysical context of star formation, a gas cloud can be submerged
in an intense electromagnetic radiation field when the nuclear reaction
has been ignited in the central protostar. In such a situation, the
large-scale gas cloud dynamics of contraction and expansion may be
regarded as quasi-isothermal. Similar conditions may be applicable
in the emergence of a planetary nebula system as well as in a certain
evolution phase for hot gas in a galaxy cluster. Although not fully
understood in details, there are other possible heating processes that
may happen in the EE portion such as micro-turbulence, flow instabilities,
shocks, wave dampings and magnetic fields etc.

With these considerations in mind, it is worthwile to note that
the isothermal assumption is a special case of the more general
polytropic approximation
(Cheng 1978; Goldreich \& Weber 1980; Bouquet et al. 1985; Yahil
1983; Suto \& Silk 1988; Maeda et al. 2002; Harada et al. 2003).
Asymptotic solutions parallel to those of $(16)-(18)$ as well as
eigensolutions across the sonic critical line can be readily
derived. Therefore, EECC self-similar solutions in the polytropic
approximation can be obtained with combined numerical and
analytical analyses. One major difference is that, in the
polytropic approximation, the asymptotic solution takes the form
of
\begin{equation}
v\rightarrow Vx^{(1-\Gamma)/(2-\Gamma)}\ ,
\end{equation}
\begin{equation}
\alpha\rightarrow Ax^{-2/(2-\Gamma)}\ ,
\end{equation}
\begin{equation}
m\rightarrow Dx^{(4-3\Gamma)/(2-\Gamma)}\ ,
\end{equation}
where $x\propto rt^{(\Gamma -2)}$ is the independent similarity
variable, $\Gamma$ is the polytropic index, and $V$, $A$ and $D$ are
three constant coefficients to the leading order. For $1<\Gamma<2$,
the radial flow speed (either outflows or inflows) vanishes as
$x\rightarrow +\infty$. For a finite radial range, the basic concept
of EECC similarity solutions advanced in this paper remains valid.
Additional similarity solution features such as shocks, subsonic
radial oscillations, central core collapse etc. are also expected.
Polytropic EECC similarity solutions should have a wider range of
astrophysical applications with, at least, one more degree of
flexibility (i.e. $\Gamma$ parameter).

Conceptually, the EECC similarity solutions provide a simple
model scenario for the concurrence of inner infalls (including
central core collapses) and outer envelope outflows (including
asymptotic winds) with the separation surface (or stagnation
point) travelling outward at a constant speed. In any
self-gravitating spherical fluid system, the emergence of an
EECC similarity phase depends upon allowed physical conditions
that can be reached via a certain set of initial/boundary
conditions of density and velocity at some time ago through other
processes that were usually not self-similar.

For applications of the EECC similarity solutions, as well as
other similarity solutions, we have noted several potential areas.
The first one is the dynamical evolution phase connecting the AGB
or post-AGB phase and the proto planetary nebula (PN) phase, where
the central core collapse to form a proto white dwarf and the
envelope wind to create various PN morphologies are expected to
happen simultaneously in a timescale of $\sim 10^{3}$ yrs (e.g.
Balick \& Frank 2002). By this process, a late-type star can
continue to lose its envelope mass while forming a compact object
at the collapsing core. After the self-similar phase, the system
continues to evolve in a detached manner with the core and
envelope separated. Shocks can be introduced either at the end
of the AGB phase (Kwok 1982, 1993) when the faster envelope expansion
during the EECC phase catches up with the slower pre-existing wind
or when infalling materials bounced back from the central compact
object. Much as Tsai \& Hsu (1995) and Shu et al. (2002) introduced
shocks to the similarity solutions in star formation and `champagne
flows' of H{\sevenrm II} regions, one can construct EECC similarity
solutions with shocks in the context of planetary nebulae. The
solutions other than the EECC ones (including their time-reversal
counterparts) have been used in star-formation as well as other
gravitational collapse and flow problems.

For astrophysical systems of even larger scales such as clusters of
galaxies (e.g. Sarazin 1988; Fabian 1994), the similarity solutions
can be valuable for understanding a certain phase of their evolution.
Specifically, if their evolution involves a self-similar phase of
central collapse (e.g. Gunn \& Gott 1972; Fillmore \& Goldreich
1984; Bertschinger 1985; Navarro et al. 1997), then isothermal
similarity solutions (17) and (18) seem to suggest two possible
classes of galaxy clusters that emit X-rays through hot gases
virialized in
gravitational potential wells, namely, those with steep gravitational
potential wells and thus extremely high X-ray core luminosities, and
those with relatively smooth and shallow gravitational potential wells
and thus normal X-ray core luminosities.

Finally, we note that the stability of similarity solutions is of
interest in various contexts. As several authors have declared (Ori
\& Piran 1988; Hanawa \& Nakayama 1997), some self-similar solutions
may not be stable when one performs a normal mode analysis. The
instability growth rates are different for various types of similarity
solutions in general. At this moment, the stability problem for the
EECC similarity solution remains open for future research. In some
cases, depending on the growth rates, an unstable EECC similarity
solution might be of interest to disrupt a self-similar phase within
proper timescales. For example, we do not expect a self-similar phase
to last forever, especially for the ephemeral phase of $\sim 10^3$ yrs
between the end of AGB phase and the PN phase (Kwok 1982, 1993; Balick
\& Frank 2002).

\section*{Acknowledgments}
This research has been supported in part by the ASCI Center for
Astrophysical Thermonuclear Flashes at the University of Chicago
under Department of Energy contract B341495, by the Special Funds
for Major State Basic Science Research Projects of China, by the
Tsinghua Center for Astrophysics, by the Collaborative Research
Fund from the National Natural Science Foundation of China (NSFC)
for Young Outstanding Overseas Chinese Scholars (NSFC 10028306) at
the National Astronomical Observatory, Chinese Academy of Sciences,
and by the Yangtze Endowment from the Ministry of Education through
the Tsinghua University. Affiliated institutions of Y.Q.L. share
the contribution.

\begin{appendix}
\section{Asymptotic Solutions}
In this Appendix, we describe and summarize asymptotic behaviours
of similarity solutions for $x\rightarrow +\infty$ and
$x\rightarrow 0^{+}$, and for Taylor expansions across the sonic
critical line.

\subsection{Asymptotic solutions for $x\rightarrow +\infty$}

As $x\rightarrow +\infty$ (i.e. at an initial instant
$t\rightarrow 0^{+}$), a sensible physical requirement is that the
radial flow speed and fluid density should remain finite. It turns
out that for a finite $v(x)$, the reduced density $\alpha(x)$
scales as $\propto x^{-2}$. The Taylor series of $v(x)$ and
$\alpha(x)$ for $x\rightarrow +\infty$ can be written as
\begin{equation}
\begin{split}
&v=v_0+v_1x^{-1}+v_2x^{-2}+v_3x^{-3}+...\ ,\\
&\alpha=a_2x^{-2}+a_3x^{-3}+a_4x^{-4}+...\ ,
\end{split}
\end{equation}
where $v_i$ ($i=0,1,2,\cdots$) and $a_j$ ($j=2, 3,\cdots$) are
series expansion coefficients. By substituting these expressions
into nonlinear ODEs (9) and (10), one can determine expansion
coefficients $v_i$ and $a_j$ by comparing powers of $x$ in order.
The first several terms are given explicitly in equation (16). As
expected, the series expansion involves two independent parameters
$V$ and $A$ as initial conditions. The parameter $V$ describes the
asymptotic radial flow speed and can take either positive or
negative values. The parameter $A$ describes the asymptotic density
distribution $\rho_0=Aa^2/(4\pi Gr^2)$, corresponding to a singular
isothermal sphere (SIS) if applied to the entire radial range. For
different values of $V$ and $A$, one can introduce various
conditions to initiate self-similar processes.

\subsection{Asymptotic solutions for $x\rightarrow 0^{+}$}

The asymptotic solution behaviours for $x\rightarrow 0^{+}$
(i.e. the final moment of $t\rightarrow +\infty$ at a given $r$)
can be derived in the same manner as $x\rightarrow +\infty$. For
the reduced velocity $v(x)$, it may either diverge (in an almost
free-fall state) or vanish for $x\rightarrow 0^{+}$. We therefore
obtain two sets of asymptotic solutions (17) and (18), keeping
only the leading-order terms. In particular, solution (17)
describes a free-fall state with a reduced core mass parameter $m_0$.

\subsection{Taylor expansions across the sonic critical line}

As the techniques of obtaining eigensolutions across the
sonic critical line have been discussed in WS and Hunter
(1986), we state the basic results below.
\subsubsection{Eigensolutions across the sonic critical line}

When the coefficients of $dv(x)/dx$ and $d\alpha(x)/dx$ vanish
on the left-hand sides of nonlinear ODEs (9) and (10), it is
necessary to impose condition (15) in order to derive finite
$dv(x)/dx$ and $d\alpha(x)/dx$. Specifically, derivatives
of $v(x)$ and $\alpha(x)$ can be computed through the
L'H\^{o}pital's rule, and two types of eigensolutions analytic
across the sonic critical line can be derived. In addition to
expressions (19) and (20), we summarize higher-order derivatives
below. For type 1 solutions, we have at the sonic critical
point $x=x_*$
\begin{equation}
\begin{split}
&\frac{d^2v}{dx^2}\bigg|_{x=x_*}=-\frac{(x_*-1)(x_*-3)}{x_*^2(x_*-3)}\ ,\\
&\frac{d^2\alpha}{dx^2}\bigg|_{x=x_*}
=\frac{2(x_*^2-8x_*+13)(x_*-3)}{x_*^3(x_*-3)}\
,\\
&\frac{d^3v}{dx^3}\bigg|_{x=x_*}
=\frac{(x_*-1)(x_*^2-3x_*+6)(x_*-3)^2}{x_*^3(x_*-3)^2(x_*-4)}\ ,\\
&\frac{d^3\alpha}{dx^3}\bigg|_{x=x_*}
=\frac{2(x_*-2)(x_*^3-19x_*^2+93x_*-147)(x_*-3)^2}{x_*^4(x_*-3)^2(x_*-4)}\
,
\end{split}
\end{equation}
and for type 2 solutions, we have
\begin{equation}
\begin{split}
&\frac{d^2v}{dx^2}\bigg|_{x=x_*}
=\frac{x_*^2-5x_*+5}{x_*^2(2x_*-3)}\
,\\
%
&\frac{d^2\alpha}{dx^2}\bigg|_{x=x_*}
=-\frac{2(x_*^2-6x_*+7)}{x_*^3(2x_*-3)}\
,\\
&\frac{d^3v}{dx^3}\bigg|_{x=x_*}
=-\frac{6x_*^5-16x_*^4-58x_*^3+266x_*^2-345x_*+150}
{x_*^3(2x_*-3)^2(3x_*-4)}\
,\\
&\frac{d^3\alpha}{dx^3}\bigg|_{x=x_*}
=\frac{6(x_*-2)(2x_*^4+2x_*^3-42x_*^2+83x_*-47)}
{x_*^4(2x_*-3)^2(3x_*-4)}\ .
\end{split}
\end{equation}
The corresponding Taylor expansions can then be constructed for
the two types of eigensolutions across the sonic critical line. We
immediately note that there are critical points where higher-order
derivatives diverge, e.g. $x_*=4$ for denominators of the
third-order derivatives of type 1 solutions in equation $(A2)$;
$x_*=3/2$ in the denominator of the second-order derivatives of
type 2 solutions in equation $(A3)$; and $x_*=3/2$ and $4/3$ in
the denominator of the third-order derivative of type 2 solution
and so forth. In general, for $x_*=1+N$ with integer $N\ge2$, the
L'H\^{o}pital's rule fails for type 1 eigensolutions (e.g. $x_*=3$
in equation $A2$); while for $x_*=1+1/N$ with integer $N\ge2$, the
L'H\^{o}pital's rule fails for type 2 eigensolutions (see section
3 of Hunter 1986). To avoid such densely distributed divergent
points as $x\rightarrow 1^{+}$ for type 2 eigensolutions, it would
be wise to use up to second-order derivatives with a properly
chosen step away from $x_*$.

\subsubsection{Classification of saddle and nodal critical points}

There exist two types of sonic critical points (Jordan \& Smith
1977), the saddle and nodal points. The classification of these
two types is based on the topological structure of solution paths
around the sonic critical point (WS). In short, the critical point
is a saddle point when the two gradients of eigensolutions $v(x)$
are of the opposite signs; while the critical point is a nodal
point when the two gradients carry the same sign (Jordan \& Smith
1977). In our problem, the critical points with $0<x_*<1$ are
saddle points, while those with $x_*>1$ are nodal points; the
special case of $x_*=1$ is degenerate.

Mathematically allowed solutions across a sonic critical point
are different for saddle and nodal points. Across a saddle
critical point, similarity solutions can pass through via the
two eigensolutions. For a node, besides the two eigensolutions,
infinitely many solutions can go cross the node along the
primary direction, that is, they all cross the node tangentially
to the eigensolution with a larger absolute value of gradient
for $v(x)$. Hunter (1986) noted that the two eigensolutions are
analytic while others are not.

Numerical integrations away from a saddle or a node along the
secondary direction [i.e. in terms of the eigensolution with
smaller absolute value of gradient for $v(x)$] are stable or
neutrally stable, respectively; while numerical integrations away
from a node along the primary direction are unstable due to
numerous path branches at the node in that direction. The key is
to choose a proper integration step away from the node and
pertinent initial values for numerical integrations along the
primary direction.

\subsubsection{Weak discontinuities}

There are two important aspects to allow weak discontinuities in
similarity solutions across a nodal critical point (Lazarus 1981;
Whitworth \& Summers 1985). The first one is to allow for those
non-analytic similarity solutions along the primary direction
across a node. The second one is to allow linear combinations (in
the local sense) of the two eigensolutions at a node, i.e. the
combined gradient is between those of the two eigensolutions (WS).
In this manner, WS have obtained their continuous bands of
similarity solutions. Weak discontinuities in similarity
solutions, as claimed by Hunter (1986), may bear little physical
significance. Mathematically, we simply point out here the
possibility of constructing WS types of band solutions from
discrete EECC as well as ECCC self-similar solutions.

\end{appendix}


\end{document}